\documentclass[aps,prx,nofootinbib,showpacs,reprint]{revtex4-1}
\usepackage{graphicx}
\usepackage{amsmath}
\usepackage{bm}
\usepackage[colorlinks=true,letterpaper=true, pdfstartview=FitV,urlcolor=blue]{hyperref}
\usepackage{color}
\usepackage{xcolor}
\usepackage{siunitx}
\usepackage{txfonts}
\usepackage{amsfonts}
\newcommand{\bG}{\bar{\Gamma}}
\newcommand{\bX}{\bar{X}}
\newcommand{\bY}{\bar{Y}}
\DeclareMathOperator{\signfunction}{sgn}
\newcommand{\sgn}[1]{\signfunction{(#1)}}
\newcommand{\bq}{\bm{q}}
\newcommand{\br}{\bm{r}}

\newcommand{\average}[1]{\langle#1\rangle}
\newcommand{\dst}{\Delta_{\text{st}}}
\newcommand{\Ec}{E_{c}}
\newcommand{\Ez}{E_{Z}}
\newcommand{\Bperp}{B_{\perp}}

\begin{document}
\title{
Spontaneous symmetry breaking and quantum Hall valley ordering on the surface of topological hexaborides
}
\author{Xiao Li}
\email{Email: lixiao@umd.edu}
\address{Condensed Matter Theory Center and Joint Quantum Institute, University of Maryland, College Park, Maryland 20742-4111, USA}
\author{Bitan Roy}
\address{Condensed Matter Theory Center and Joint Quantum Institute, University of Maryland, College Park, Maryland 20742-4111, USA}
\author{S. Das Sarma}
\address{Condensed Matter Theory Center and Joint Quantum Institute, University of Maryland, College Park, Maryland 20742-4111, USA}

\date{\today}

\begin{abstract}
A number of strongly correlated heavy fermion compounds, such as samarium (Sm), ytterbium (Yb), plutonium (Pu) hexaboride, are predicted to become topological insulators at low temperatures. These systems support massless Dirac fermions near certain (three) points of the surface Brillouin zone, hereafter referred to as the valleys. In strong perpendicular magnetic fields, the conical Dirac dispersions of these surface states quench onto three sets of Landau levels and we predict various possible hierarchies of incompressible quantum Hall states on the surface of hexaborides. In addition, we address the effects of strong electron-electron interaction within the surface zeroth Landau levels. Specifically, we show that depending on the relative strength of the long-range (Coulomb-type) and the finite-range (Hubbard-type) interactions the ground state can display either a valley-polarized or a valley-coherent distribution of electronic density. We also show that the transition between two valley-polarized states is always discontinuous, while that between a valley-polarized and a valley-coherent phase is continuous. The Zeeman splitting and/or an applied uniaxial strain on the surface can drive the system through various quantum phase transitions and place it in different broken-symmetry phases. Application of uniaxial strain is also shown to considerably modify the precise sequence of quantum Hall states. We also highlight the role of topology in determining the broken symmetry phases, disorder on the surface of topological hexaborides in strong magnetic fields.   
\end{abstract}

\maketitle

\section{Introduction}

Two-dimensional semiconductor heterostructures, such as those prepared in GaAs quantum wells and Si metal-oxide-semiconductor field-effect transistors (MOSFETs), offer unique opportunities to study many novel properties of low-dimensional electronic systems. Possibly the most celebrated example is the quantum Hall effect~\cite{prange,das2008perspectives}, where precise quantization of the Hall conductance is observed at low temperatures in the presence of a strong perpendicular magnetic field. The quantum Hall states represent the first example of topological states of matter [time-reversal symmetry breaking due to the presence of an external magnetic field, but protected by U(1) charge conservation symmetry] with a bulk gap and one-dimensional chiral edge modes at the boundary. The ground state topology of integer quantum Hall states is captured by the Thouless-Kohmoto-Nightingale-de Nijs (TKNN) invariant or the first Chern number~\cite{TKNN1982}.

In the past decade, our notion of topological states of matter has been vastly expanded. It began with the generalization of topological vacua protected by time-reversal invariance, which led to the discovery of two- and three-dimensional topological insulators and superconductors~\cite{Hasan:2010Review,Qi:2011Review, volovik2009universe,DasSarma2015:MfReview}. In fact, almost immediately after the theoretical proposal, Bi$_2$Se$_3$ was established as the first prototype candidate for three-dimensional strong $Z_2$ topological insulator~\cite{Hsieh:2008TI-ARPES,Xia:2009TI-ARPES}. These developments stimulated a surge of theoretical works, aimed at extracting topological information from various quantum phases of matter~\cite{altland-zirnbauer, rahulroy, Schnyder2008:Classification,Kitaev2009:Eightfold}.

From the perspective of quantum field theory, a topologically nontrivial ground state can be described as a $\theta$-vacuum, and different values of $\theta$ correspond to topologically distinct phases. One well studied example of such a topological quantum field theory is the Chern-Simons theory in two dimensions, where the coefficient of the $\theta$-term dictates the Hall conductivity in the system~\cite{wen2007quantum,fradkin2013fieldtheory}. 
In three dimensions, the corresponding topological field theory is tied with the magneto-electric effect, captured by an axion term $\theta_{\text{ax}}{\mathbf E} \cdot {\mathbf B}$~\cite{Qi:2008FieldTheory}. The axion angle for a time-reversal and inversion symmetric strong $Z_2$ topological insulator is $\theta_{\text{ax}}=(2n-1) \pi$, where $n$ is an integer \footnote{We note that when the ground state has a $D$-fold degeneracy, the axion angle is allowed to take fractional values $\theta_{\text{ax}} \to \theta_{\text{ax}}/D$~\cite{Senthil}. However, we do not delve into such a situation.}. A nontrivial axion angle (modulo $2 \pi$) leaves its signature in the bulk, where the insulation through the band inversion occurs at $2n-1$ high-symmetry time-reversal invariant momentum (TRIM) points of the Brillouin zone. Through the bulk-boundary correspondence, the surface hosts $2n-1$ copies of two-component massless Dirac fermions. Angle-resolved-photo-emission-spectroscopy (ARPES)~\cite{Hsieh:2008TI-ARPES,Xia:2009TI-ARPES,Hsieh:2009TI,Hsieh:2009TIARPRES,Roushan:2009TI-ARPRES}, scanning tunneling microscopy (STM)~\cite{Hor:2009TI-STM} and quantum oscillation measurements~\cite{Analytis:2010TI} have confirmed the existence of such topologically protected Dirac cones on the surface of various topological insulators. In Bi$_2$Se$_3$, for example, the bulk band inversion occurs only at the $\Gamma$ point. As a result, the surface state is comprised of only a single copy of massless helical Dirac cone, carrying a nontrivial Berry phase $\pi$.

\begin{figure}[!]
\includegraphics[scale = 0.55]{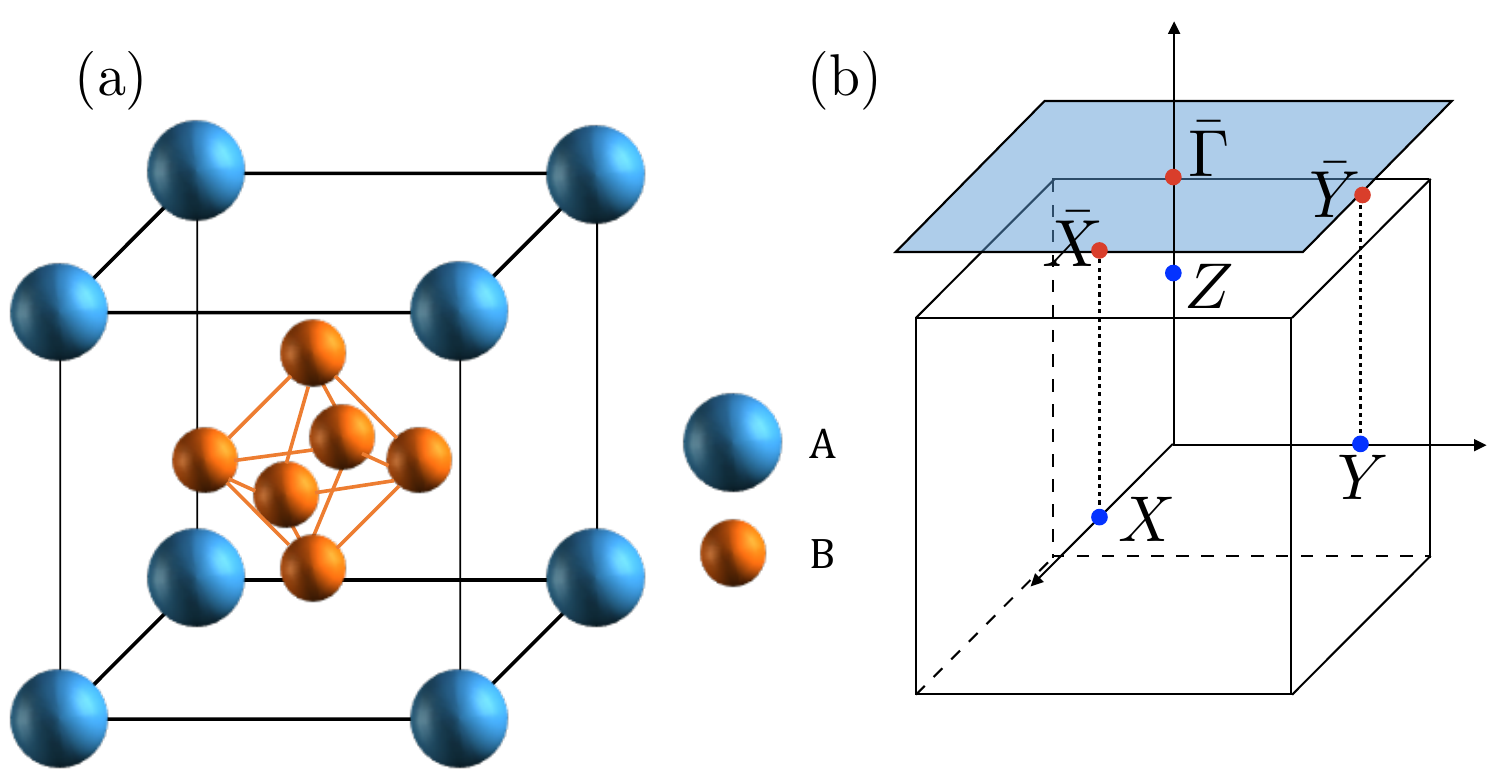}
\caption{(a) Crystal structure of topological hexaborides (AB$_6$). Here, B represents boron atoms, six of which form a cage at the center of the cubic lattice, and A$=$Sm, Yb, Pu, for example. (b) Bulk Brillouin zone for AB$_{6}$ and its projection on (001) surface. The bulk band inversion takes place at three high-symmetry points $X = (\pi, 0, 0)$, $Y = (0, \pi, 0)$, and $Z = (0, 0, \pi)$ (marked by blue dots). As a result, the $(001)$ surface hosts three Dirac cones (marked by red dots). \label{Fig:Lattice} }
\end{figure}

Recently a family of hexaboride compounds, such as SmB$_6$~\cite{Dzero-original, Dzero-prb, Miyazaki2012:SmB6, Alexandrov2013:SmB6,Yee2013:SmB6,Neupane2013:SmB6,Jiang2013:SmB6,Xu2013:SmB6,Frantzeskakis2013:SmB6,Denlinger2013:SmB6,Min2014:SmB6,Legner2014:SmB6,Fuhrman:2015SmB6}, YbB$_6$~\cite{Weng:2014YbB6,Xia:2014YbB6ARPRES,Neupane:2015YbB6,Xu:2014ARPRES} and PuB$_6$~\cite{Deng:2013PuB6}, together named \emph{topological hexaborides} (THBs),~\footnote{SmB$_6$, PuB$_6$ are also referred to as ``topological Kondo insulators''} have been predicted to be topological insulators at low temperatures. As shown in Fig.~\ref{Fig:Lattice}, these systems have cubic unit cells, with six boron atoms forming a cage at its center, while one $A$ atom ($A$=Sm, Yb, Pu) is shared by eight corners. 
In the cubic environment the band inversion takes place at three $X$ points in the bulk Brillouin zone, and is labeled as $T-p3(4)_X$ in the space group classification of topological insulators~\cite{juricic}. Consequently each surface accommodates three copies of gapless Dirac cones. On the high symmetry (001) surface, for example, the Dirac points are located at $\bar{\Gamma}=(0,0)$, $\bar{X}=(\pi,0)$, and $\bar{Y}=(0,\pi)$, as shown in Figs.~\ref{Fig:Lattice}(b) and \ref{Fig:LL}(a). Therefore, THBs can be considered as topological insulators with an axion angle $\theta_{\text{ax}}=3\pi$, and they belong to the same class AII in ten-fold way of classification as Bi$_2$Se$_3$~\cite{Schnyder2008:Classification}. Furthermore, the chemical potential in THBs resides within the bulk insulating gap, offering rare examples of \emph{true} topological insulators, unlike the situation in typical semiconductor topological insulators, such as Bi$_2$Se$_3$ which inevitably tend to have considerable bulk conduction due to inherent doping induced by defects. The insulating bulk nature of THBs, along with growing experimental evidence for a two-dimensional (2D) surface conducting layer, has led to much activities in these systems.

In contrast to conventional (weakly correlated) topological insulators such as Bi$_2$Se$_3$, the bulk gap in THBs stems from strong interaction-driven hybridization between the Kramers degenerate bands with opposite parities. For example, the hybridization predominantly takes place between $d$ and $f$ orbitals in SmB$_6$ and PuB$_6$, and between $p$ orbitals in YbB$_6$. As a result, strong electronic interactions, aside from giving rise to topologically nontrivial bulk insulators, may as well lead to interesting many-body phenomena on the surface~\cite{Roy:2014vv,Efimkin:2014,Chen2014:OpticalTKI,Nikolic2014:Kondo,Baruselli2014:STM-SmB6,Triola2015:Manybody}. Therefore when placed in strong magnetic fields, the surface of THBs turns into a fertile ground for exploring the interplay of topology, electronic interactions, and the quantum Hall phenomena, both theoretically and experimentally. Recent success in experimentally observing surface quantum Hall states in conventional topological insulators~\cite{Qikun:2010PRL, Xu:2014NatPhys} and the presence of strong residual interaction on the surface of THBs, motivates this work. We here study various possible broken-symmetry quantum Hall phases in two-dimensional metallic surface of THBs.

Recent experiments suggest that certain hexaboride systems, in particular SmB$_6$ and YbB$_6$, can become topological insulators (i.e., with a bulk gap and an odd number of surface Dirac cones) at low temperatures. 
For example, recent ARPES experiments have confirmed the existence of three Dirac cones around $\bG$, $\bX$, $\bY$ points of the surface Brillouin zone, and the spin-resolved ARPES is strongly suggestive of their spin-momentum locking~\cite{Frantzeskakis2013:SmB6,Jiang2013:SmB6,Neupane2013:SmB6,Denlinger2013:SmB6,Xu2013:SmB6,Min2014:SmB6} (the hallmark signature of surface states in strong $Z_2$ topological insulators). ARPES has further exposed that while the Dirac cone at $\bG$ is circular, those near $\bX$ and $\bY$ are elliptic. Signature of surface metallic states has also been observed in STM measurements~\cite{Yee2013:SmB6}. In addition, De Haas-van Alphen effect~\cite{Li:2014hk} has found traces for two distinct sets of Dirac-type Landau levels up to $45$ Tesla (one from the $\bG$ pocket, and other one from the $\bX$ and the $\bY$ pockets together). An infinite-field ($B \rightarrow\infty$) extrapolation from these data reveals a $-\pi$ Berry phase for each such Dirac cone. These observations are in agreement with recent theoretical analyses that predict a generic off-set between the energies of the Dirac points at $\bG$ ($E_{\bG}$) and $\bX$, $\bY$ ($E_{\bX/\bY}$) points. Such an outcome can also be anticipated based on a symmetry analysis: while the underlying cubic symmetry mandates $E_{\bX}=E_{\bY}$ on the (001) surface, $E_{\bG}$ remains unconstrained and typically $|E_{\bG}-E_{\bX/\bY}|=\Delta_0 \sim 2-10$ meV~\cite{Roy:2014dg, XDai:2015}. Magnetoresistance measurements in a \emph{wedge}-shaped SmB$_6$ sample showed that transport at low temperatures ($T<\SI{10}{K}$) is independent of the sample thickness and dominated by the surface states~\cite{kim2013surface}. These experimental observations, when combined with recent tight-binding~\cite{Roy:2014dg} and first-principles~\cite{XDai:2015} calculations, enable us to establish a minimal model [see Eq.~(\ref{Eq:H0})] to study the effects of strong magnetic field and electron-electron interaction on the surface of THBs.

In this work we study the interplay of external magnetic field and strong Coulomb repulsion on the surface of THBs. Following, we promote the central results of our theoretical study and the organization principle for the rest of the paper. 

(i) In Sec.~\ref{Section:LL}, we account for the effects of strong magnetic field on the surface of THBs and present the spectrum of surface Landau levels. Using realistic values of effective band parameters, we show that the crossing between the zeroth Landau levels and those at finite energies is avoided for magnetic fields beyond $B_{\ast}\sim \SI{1}{T}$, see Fig.~\ref{Fig:LL}(b). Perturbations from Zeeman coupling and/or uniaxial strain do not change this result quantitatively. Furthermore, we show that depending on the relative position and the anisotropy of these surface Dirac cones, a number of distinct hierarchies of quantum Hall states can be observed in these systems.   

(ii) Within the framework of a zeroth Landau level (LL) approximation (i.e., neglecting LL mixing which are negligible at high magnetic fields), we develop the mean-field Hartree-Fock theory for interacting surface states in Sec.~\ref{Section:Interaction}. Both long-range (Coulomb-type) and short-range (Hubbard-type) interactions are accounted for in the theory, which should be exact in the limit of a very large inter-LL gap, since the Hartree-Fock theory is a systematic expansion in the ratio of interaction to the LL separation energy. This section is slightly technical and readers interested in the results may choose to skip the discussion. 

(iii) The phase diagrams of interacting surface states in strong magnetic fields are discussed in Sec.~\ref{Section:GroundStates}. We show that for long-range Coulomb interactions the electronic density can only display valley polarizations. Due to the anisotropic cyclotron orbits near $\bX$ and $\bY$ valleys, the long-range interaction prefers a $\bG$ polarized state [see Fig.~\ref{Fig:PhaseCoulomb}(a)], which does not break any microscopic symmetry. On the other hand, the $\bX$ or $\bY$-valley polarized states break the discrete rotational symmetry of the underlying lattice, and correspond to the \emph{quantum Hall Ising nematic order}. When finite-range or backscattering interactions are included, a valley-coherent phase can also be realized [see Fig.~\ref{Fig:PhaseCoulomb}(b)]. Finally, we show that various single-particle perturbations, such as the Zeeman coupling and the uniaxial strain serve as useful tuning parameters to drive the system through various quantum phase transitions and place the system in different broken-symmetry phases (see Figs.~\ref{Fig:PhaseZeeman} and \ref{Fig:PhaseStrain}).

(iv) We summarize our findings and present concluding remarks in Sec.~\ref{Section:Discussions}. 
In addition, we propose experimental signatures, such as the measurement of longitudinal conductivity ($R_{jj}$) and the scaling of $\Delta R=|R_{xx}-R_{yy}|$ with magnetic field at finite temperatures that can pin the exact nature of the broken symmetry states within the zeroth Landau level. The role of disorder in the ordered phases, and the importance of underlying topology in determining the ground states are also highlighted in this section.

\section{Surface states: external magnetic and strain fields\label{Section:LL}}

\begin{figure}[!]
\includegraphics[scale = 0.6]{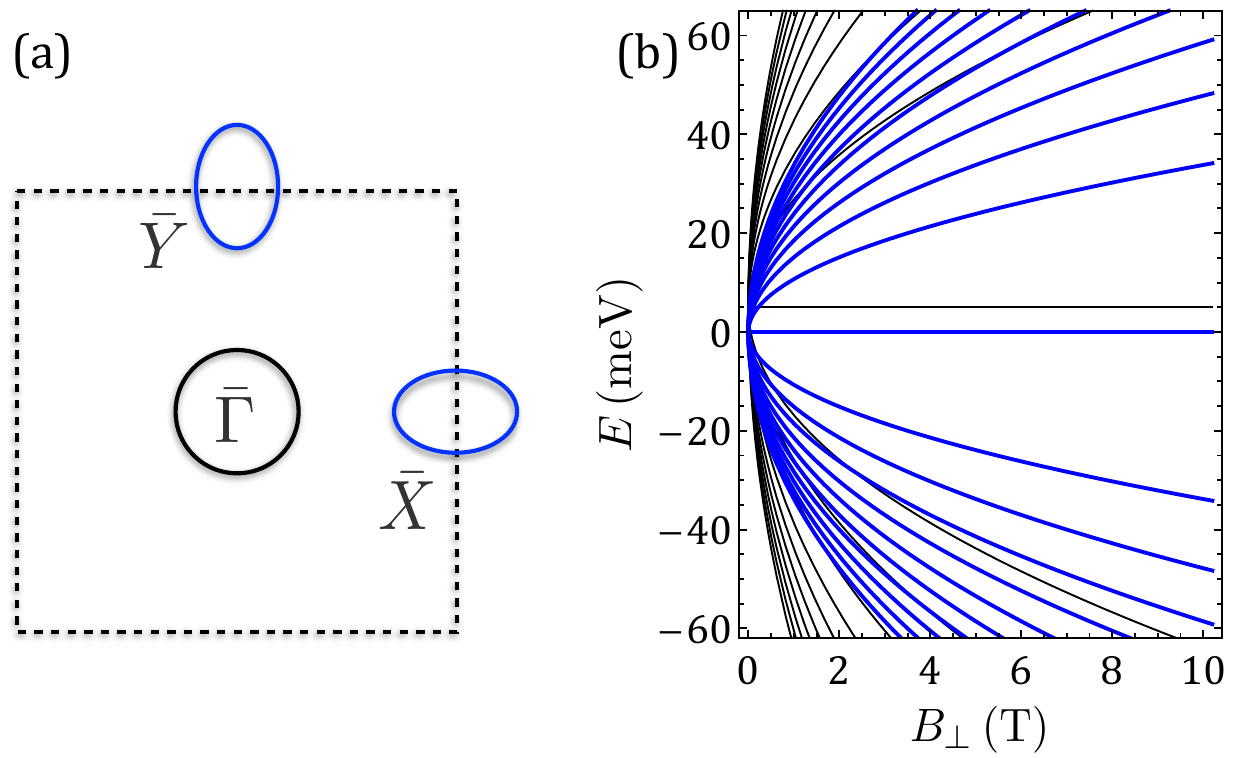}
\caption{(a) Illustration of three electron pockets, and (b) the Landau level spectrum (for $|N| \leq 10$) on the (001) surface of THBs. The black (blue) curves represent the Landau levels from the $\bG$ ($\bX$ or $\bY$) valley. The energy offset between these two sets of Landau levels is $\Delta_0=\SI{5}{meV}$, the anisotropy factor is $\delta=0.2$, {and the Fermi velocities are chosen as $v_{\bG}= 6.0\times10^{5}$\,{m/s} and $v_{\bX/\bY}= 3.0\times10^{5}$\,{m/s}.} \label{Fig:LL} }
\end{figure}

To set a context for our study, we start by introducing an effective single-particle description on the surface of THBs. We focus on the high-symmetry $(001)$ surface, which hosts three copies of massless Dirac fermionic excitations at $\bG$, $\bX$, and $\bY$ points of the 2D surface Brillouin zone, hereafter referred to as \emph{valleys}, as shown in Fig.~\ref{Fig:LL}(a). The effective Hamiltonian for these surface states is given by~\cite{Roy:2014vv, Roy:2014dg}
\begin{align}
	H_{\lambda} &= {\hbar v_{\lambda} \left[ (1+\delta_{\lambda})k_x\sigma_x-(1-\delta_{\lambda})k_y\sigma_y \right]} \notag\\ &\quad + \delta H^Z_\lambda + \delta H^{\text{str}}_\lambda, \label{Eq:H0}
\end{align}
where $\lambda=1, 2, 3$ corresponds to the $\bG$, $\bX$, and $\bY$ valley, respectively. 
{ The Fermi velocity of the surface Dirac fermions is taken to be $v_{\bG}= 6.0\times10^{5}$\,{m/s} and $v_{\bX/\bY}= 3.0\times10^{5}$\,{m/s} for our numerical work, in accordance with recent quantum oscillation experiments}~\cite{Li:2014hk}. 
{However, note that quantum oscillation experiments cannot pin the Fermi velocity associated with the Dirac cone at a given valley. Thus, we combine the information from quantum oscillation experiments with ARPES data~\cite{Frantzeskakis2013:SmB6,Jiang2013:SmB6,Neupane2013:SmB6,Denlinger2013:SmB6,Xu2013:SmB6,Min2014:SmB6}, which suggests that the surface state near $\Gamma$ point has a higher Fermi velocity.}

The parameter $\delta_{\lambda}$ characterizes the anisotropic Fermi velocity for the surface states near each valley. For a pristine surface, we set $\delta_{\bG}$\,$=$\,$0$ (representing a circular Fermi surface near the $\bG$ valley) and $\delta_{\bX} = -\delta_{\bY} =\delta$ (accounting for elliptic Fermi surfaces near $\bX$ and $\bY$ valleys). In addition, the surface states at $\bX$ and $\bY$ valleys are energetically degenerate due to a four-fold rotational symmetry. However, the surface state at the $\bG$ valley in general has a different energy. The energy difference between the $\bG$ and $\bX/\bY$ valleys can be tuned independently, for example, through band bending effects~\cite{Roy:2014dg}. To capture this energy difference, we introduce a constant shift $\Delta_0$ to $H_{\bG}$. The last two terms $\delta H^Z_\lambda$ and $\delta H^{\text{str}}_\alpha$ arise from the Zeeman coupling and an external uniaxial strain, respectively, and will be discussed later. 

\subsection{Surface Landau levels}

In the presence of a uniform perpendicular magnetic field $B$, the surface states support three sets of Landau levels. For now we neglect the effects of Zeeman coupling and external strain. 
The Landau level spectrum can be found by taking the Peierls substitution $\hbar \bm{k} \to \bm{\pi}=\hbar \bm{k}+e\bm{A}$. In the Landau gauge the vector potential is given by $\bm{A} = (0, B x)$, from which we define three pairs of creation and annihilation operators~\cite{Li:2013tx, Xiao:2015TCI-QH}
\begin{align}
	a_{\lambda} = \dfrac{\ell}{\sqrt{2}\hbar}\bigl(\alpha_{\lambda}\pi_{x} - i\frac{\pi_{y}}{\alpha_{\lambda}}\bigr), \;
	a_{\lambda}^{\dagger} = \dfrac{\ell}{\sqrt{2}\hbar}\bigl(\alpha_{\lambda}\pi_{x} + i\frac{\pi_{y}}{\alpha_{\lambda}}\bigr), 
\end{align}
where $\ell=\sqrt{\hbar/e B} $ is the magnetic length. The factor $\alpha_{\lambda}$ accounts for the anisotropy of the Fermi surface at each valley, with $\alpha_{\bX} \!=\! \sqrt{(1+\delta)/(1-\delta)}\!=\!\alpha_{\bY}^{-1}$, and $\alpha_{\bG} \!=\! 1$. 
In terms of these creation and annihilation operators, the surface Hamiltonian~\eqref{Eq:H0} takes the form
\begin{equation}
H_\lambda= \frac{\sqrt{2} \hbar v_{\lambda}}{\ell} \left( \begin{array}{cc}
0 & a^\dagger_\lambda \\
a_\lambda & 0
\end{array}
\right) + \; \Delta_0 \delta_{\lambda,1},
\end{equation}
for $\lambda=1,2,3$. The Landau level spectrum near each valley can then be readily obtained, yielding   
\begin{align}
	E_{N}(\bG) &= \Delta_0 +\sgn{N}\sqrt{2e\Bperp\hbar v_{\bG}^2 |N|}, \notag\\
	E_{N}(\bX/\bY) &= \sgn{N}\sqrt{2e\Bperp\hbar v_{\bX}^2 |N|(1-\delta^2)}, 
\end{align}
where $N=0,1,2, \cdots,$ is the Landau level index. The Landau level spectrum for $\delta=0.2$ and $\Delta_0 = \SI{5}{meV}$ is shown in Fig.~\ref{Fig:LL}(b). {With these parameters the crossing between the zeroth Landau levels and the remaining ones is avoided for fields beyond a critical value $B_\ast$ defined as}
\begin{align}
	{B_\ast=\dfrac{\Delta_0^2}{2e\hbar v_{\bX}^2(1-\delta^2)}\approx \SI{0.65}{T}. }
\end{align}
The absence of Landau level crossings at large fields will become crucial when we account for electron-electron interactions within the zeroth Landau level. 

The wave function for the $N^{th}$ Landau levels at each valley is given by 
\begin{align}
	\phi_{NX}^{(\lambda)}(\br) = \dfrac{e^{iXy/\ell^2}}{\sqrt{2L_y}}
	\begin{pmatrix} 
	\sqrt{1+\delta_{N,0}} \,\phi_{|N|}^{(\lambda)}(x-X) \\
	(1-\delta_{N,0})\sgn{N} \,\phi_{|N|-1}^{(\lambda)}(x-X)
	\end{pmatrix}, \notag
\end{align}
where $L_y$ is the linear dimension of $(001)$ surface in the $y$-direction, $X$ is the guiding center coordinate, and $\phi_{N}^{(\lambda)}(x)$ is the harmonic oscillator wave-function, 
\begin{equation}  
\phi_{N}^{(\lambda)}(x)= \frac{e^{-x^2/2\alpha_{\lambda}^2\ell^2}}{(2^{N}N!\sqrt{\pi}\alpha_{\lambda}\ell)^{1/2}} \: H_{N}(x/\alpha_{\lambda}\ell), 
\end{equation} 
where $H_N$ is the Hermite polynomial of order $N$. Note that the wave function for the zeroth Landau level near each valley is spin-polarized. 
Therefore, on two opposite surfaces of a topological insulator the zeroth Landau levels are projected onto identical spin components. In contrast, the components of the spinor basis in graphene are constituted by the pseudospin (sublattice) degrees of freedom, instead of the real spin. As a result, in graphene the zeroth Landau levels near two opposite valleys reside on complementary sublattices for each spin projection. 

{Although we here assumed three valleys to be disconnected, the bands always get connected to each other at high energy. However, such connectivity of the bands from three valleys does not spoil the existence of zeroth Landau levels, as the chiral nature of the dispersive surface states mandates the existence of three fold zeroth Landau level. The robustness of zeroth Landau level in the presence of a magnetic field of arbitrary profile (as long as a finite flux gets trapped by the system) stems from the \emph{Aharonov-Casher index theorem}~\cite{aharonov-casher}, which remains operative even on a lattice~\cite{roy-catalysis}.}

{So far we have assumed that surface Dirac points reside within the bulk insulating gap, which may not be the generic situation in reality. Within the framework of an effective description for TIs, the bulk gap can be considered as a continuous chiral symmetry breaking (but parity and time-reversal symmetry preserving) mass for three dimensional Dirac fermions. With the application of external magnetic fields, the amplitude for the chiral symmetry breaking mass increases monotonically~\cite{roy-sau-catalysis}, while the surface zeroth Landau level continue to rest at the Dirac point. Therefore, if the surface Dirac points are buried within few meV of bulk valence band (in the absence of magnetic field), the surface zeroth LLs can be pulled out from the sea of three dimensional dispersive LLs, and our following discussion remains justified in the strong magnetic fields limit. It should, however, be noted that when surface Dirac points gets immersed deep inside the bulk valence band [say $\SI{\sim100}{meV}$, although such situation is quite pathological in the context of all known three dimensional topological insulators (TIs)], even strong magnetic field ($\SI{\sim30}{T}$) may not be sufficient to place the surface zeroth LL within the bulk gap. But, we do not delve into such unrealistic situation in this work. }

\subsection{Quantum Hall states in topological hexaborides}

\begin{figure}[!]
\includegraphics[scale = 0.51]{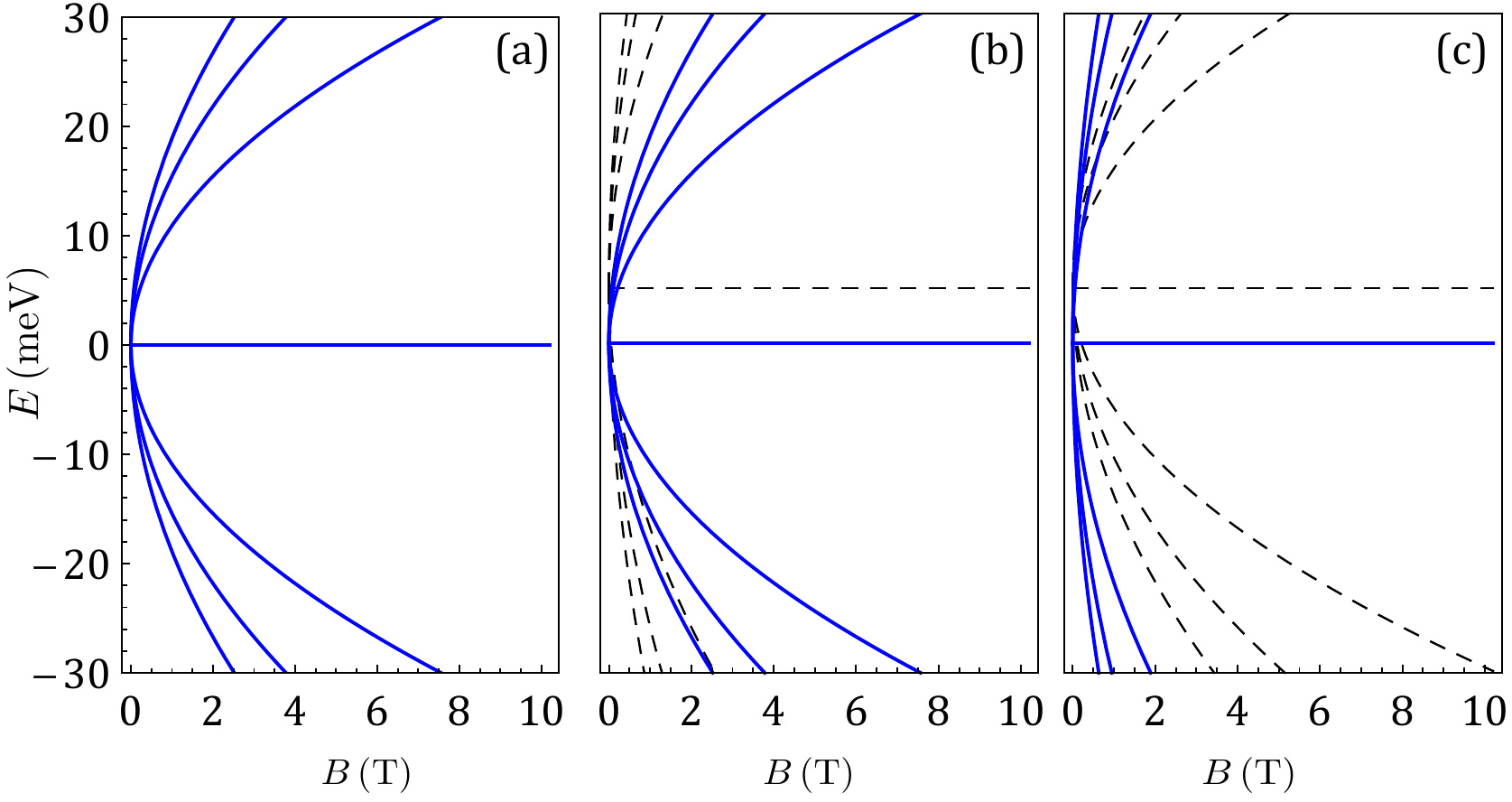}
\caption{{Lowest few Landau levels ($|N|\leq 3$) for various possible effective band parameters for the surface states.
(a) $\Delta_0 = 0$, $v_{\bG} = v_{\bX} = 3\times 10^{5}\,\text{m/s}$, and $\delta=0$; (b) $\Delta_0 = \SI{5}{meV}$, $v_{\bX} = v_{\bG}/2 = 3\times 10^{5}\,\text{m/s}$, and $\delta=0.2$; (c) $\Delta_0 = \SI{5}{meV}$, $v_{\bG} = v_{\bX}/2 = 3\times 10^{5}\,\text{m/s}$, and $\delta=0.2$.} The solid and dashed lines represent Landau levels near the $\bG$ and $\bX$/$\bY$ valleys, respectively. As a result, the blue lines are all doubly degenerate, while the black ones are non-degenerate. \label{Fig:Filling} }
\end{figure}

Having described the Landau level spectrum for the surface states, we can now discuss the possible quantum Hall states in this system. As we show in the following, depending on the relative position of the Dirac points and their Fermi velocities, different hierarchies of quantum Hall states can be realized on the surface of THBs. Thus, the sequence of Hall plateaus can provide valuable information about surface states of THBs. Remember that both top and bottom surfaces undergo Landau level quantization in the presence of a perpendicular magnetic field. Hence, all Hall plateaus we propose in the following should be augmented by an \emph{additional factor} of $2$. 
In addition, while the top and bottom surfaces host Landau levels and thus chiral edge states, the four side surfaces accommodate nonchiral helical states (assuming the sample to be cubic). These states can destroy the perfect quantization of Hall conductivity~\cite{Lee2009:Corbino, Vafek2011:QHE-TI, larsfritz}. However, recently clear signatures of Hall plateaus have been observed on the surface of Bi$_2$Se$_3$~\cite{Xu:2014NatPhys}. Therefore, we expect that our proposed Hall states can also be observed on the surface of SmB$_6$ and YbB$_6$, especially given that Dirac-type Landau levels have been observed for the surface states of SmB$_6$~\cite{Li:2014hk}.

Let us begin the discussion with a very simple situation: 
(i) $\Delta_0 = 0$, $v_{\bG} = v_{\bX}$ and $\delta=0$, for which the Landau level spectrum is shown in Fig.~\ref{Fig:Filling}(a). Surface states then host three-fold degenerate Landau levels and the Hall plateaus appear at fillings $|\nu|=3 \left( \frac{1}{2} + N \right)$, where $N=0,1,\dots$. 
{
(ii) Next we consider a realistic scenario with $\Delta_0 = \SI{5}{meV}$, $v_{\bX} = v_{\bG}/2 = 3\times 10^{5}\,\text{m/s}$, and $\delta=0.2$, for which the Landau level spectrum is shown in Fig.~\ref{Fig:Filling}(b). In this case the degeneracy of Landau levels between the $\bG$ valley and the $\bX/\bY$ valley is lifted, and the lowest few Hall plateau sequence will be $\nu = \dots, -\frac{7}{2}, -\frac{3}{2}, \frac{1}{2}, \frac{3}{2}, \frac{7}{2}, \dots$. 
(iii) Finally, we focus on the situation when the $\bG$ valley electrons have smaller Fermi velocities, i.e., $\Delta_0 = \SI{5}{meV}$, $v_{\bG} = v_{\bX}/2 = 3\times 10^{5}\,\text{m/s}$, and $\delta=0.2$. As shown in Fig.~\ref{Fig:Filling}(c), the lowest few Hall plateau will show up for filling factors at $\nu = \dots, -\frac{7}{2}, -\frac{5}{2}, -\frac{3}{2}, \frac{1}{2}, \frac{3}{2}, \frac{5}{2}, \frac{7}{2}, \dots$. 
Thus, from the sequence of Hall plateaus, one can pin the relative location of the Dirac cones and their Fermi velocities. 
}

{Throughout this paper, we will assume that the sample of TKIs is thick, and the system is far from the transition point between the topological and the trivial insulating phases, so that the overlap among the states residing on the top and the bottom surfaces is extremely weak. As a result, the total contribution to the Hall conductivity from the top and the bottom surfaces simply adds up. Thus, our analysis belongs to the opposite limit from the one in quantum Hall bilayer system, in which the interlayer tunneling is strong. By contrast, in the thin film limit the interlayer tunneling is strong and the analysis will closely resemble to the one for bilayer Hall systems. However, we do not discuss the thin film limit in this work, rather leave it as a subject for future investigation. }

\begin{figure}[!]
\includegraphics[scale=0.6]{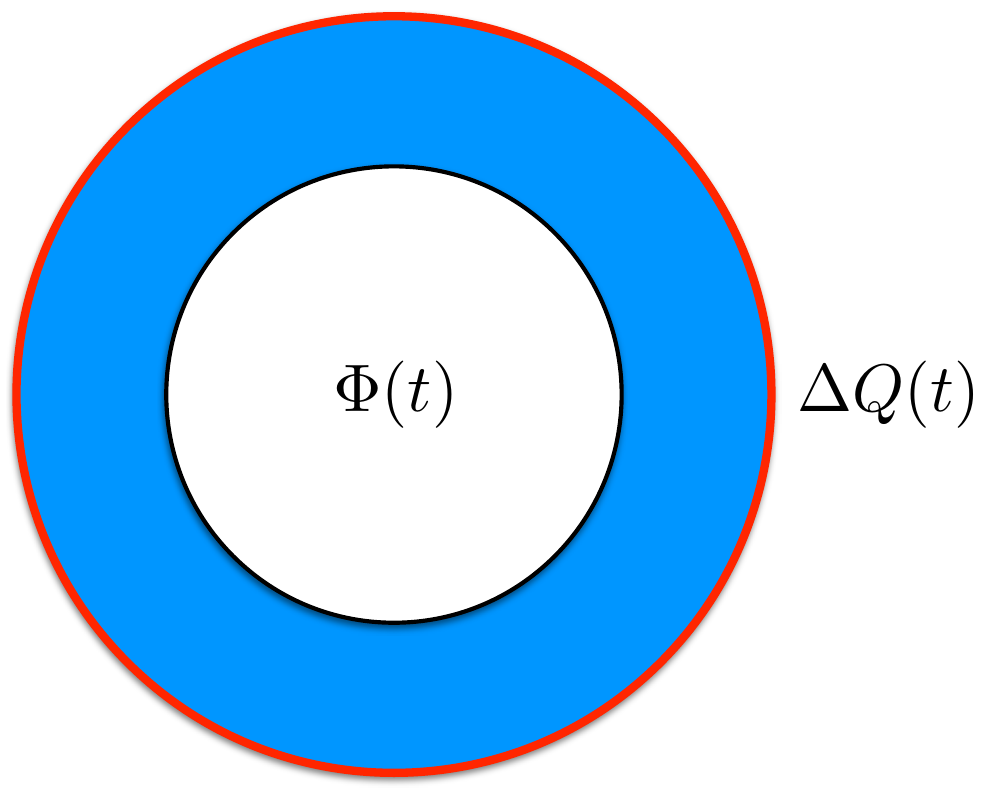}
\caption{Corbino-shaped THBs. A time-dependent flux $\Phi(t)$ pierces through the hollow part of the cylinder, and $\Delta Q(t)$ is charge accumulated on the outer surface. \label{Fig:corbino}}
\end{figure}

As mentioned earlier, the presence of helical states on the side surfaces can, in principle, ruin the perfect quantization of Hall conductivity. As an alternative, experiments in a \emph{Corbino-shaped} geometry can be carried out to access various Hall states. As shown in Fig.~\ref{Fig:corbino}, when a time dependent flux $\Phi(t)$ pierces through the hollow part of the cylindrical shaped system, there is a net flow of charge from the inner to the outer surface or vice-versa, depending on the direction of the applied magnetic field. The net charge accumulation on one side of the cylinder $\Delta Q(t)$ is proportional to the Hall conductivity ($\sigma_{xy}$) according to $\Delta Q(t)= \sigma_{xy} \: \Delta \Phi(t)$, where $\Delta \Phi(t)$ is the change of flux over time $t$. Such an experiment has been carried out successfully in regular 2D electron gas, and the signature of various integer as well as fractional Hall states has been observed~\cite{dogopolov}. We believe that similar experiments can also be carries out in topological insulators to explore the quantum Hall phenomena on its surface. 

The above proposal for quantum Hall measurement in TKIs with Corbino geometry can be considered as a generalization of Laughlin's analogous proposal~\cite{Laughlin1981:QHE} for two dimensional electron gas systems, which follows a gauge invariance argument that the accumulation of charge on one surface (inner or outer depending on the direction of magnetic field) of the Corbino disk is proportional to the Hall conductivity in the system. The charge accumulation takes place only due to the topologically protected chiral edge states, while the helical states from the side surfaces do not contribute in this process~\cite{Vafek2011:QHE-TI}. { With the recent success of observing quantum Hall states on the surface of TIs (Bi$_2$Se$_3$)~\cite{Xu:2014NatPhys}, implementation of corbino geometry to observe the Hall states may not be necessary to demonstrate similar phenomena in TKIs. However, it remains as an interesting proposal to further pin the topological nature of the Hall states on the surface of TKIs (in particular the fractional Hall states).    
}

\subsection{Zeeman splitting and uniaxial strain \label{Section:ZeemanStrain}}

Next we explore the effects of various single-particle perturbations, such as the Zeeman coupling and external strain on the surface states. We argue that Zeeman coupling introduces a \emph{mass} for these surface Dirac fermions. In addition, when a uniaxial strain is applied on the surface along a certain direction ($\bG$-$\bX$ or $\bG$-$\bY$), the Ising-type symmetry between the $\bX$ and $\bY$ valleys can be lifted.

The Zeeman coupling is given by  
\begin{align}
	\delta H^{Z}_{\lambda} = m_{\lambda} \sigma_z, \label{Eq:Zeeman}
\end{align}
where $m_\lambda = \frac{1}{2}g_{\lambda}\mu_B B$, and $g_{\lambda}$ is the effective $g$-factor for electrons at valley $\lambda$. This perturbation shifts the zeroth Landau level to energy $m_\lambda$, and places all other Landau levels ($|N| \geq 1$) at energies $\pm E_N(\lambda)$ to $\pm \sqrt{E^2_N(\lambda)+m^2_\lambda}$. Such Dirac mass can be further enhanced due to the residual electronic interactions on the surface. 
Hence, even if the bare $g$-factors in THBs may be small, it can be substantially renormalized and enhanced by electron-electron interaction, making the Dirac mass rather large.

In order to preserve the symmetry between $\bX$ and $\bY$ valleys, we set $g_{\bX} = g_{\bY} \equiv g_1$. However, a different $g$-factor for the $\bG$ valley carriers, $g_{\bG}\equiv g_{0}$, is permissible~\cite{Zhang:2013jc}, which can be justified in the following way. The effective $g$-factor for the surface states is set by the overlap between the wave functions of two hybridizing bands with opposite parities and different $g$-factors on the surface~\cite{modeTI-Zhang}. Since, such overlap at $\bX$ and $\bY$ points are constrained to be equal due to the underlying cubic symmetry in the bulk, implying $g_{\bX} \equiv g_{\bY}$, but generically different from that at $\bG$ point, and thus $g_{\bG} \neq g_1$, in general. The difference between the $g$-factors can also be enhanced by band-bending effects near the surface~\cite{Roy:2014vv}.

The Zeeman coupling therefore causes a field-dependent energy difference between the zeroth Landau level at the $\bG$ and $\bX/\bY$ valleys, given by 
\begin{align}
	\Delta_0^Z \equiv E_{\bG}(N=0) - E_{\bX} (N=0) = \Delta_{0} + \frac{1}{2} g_\text{eff} \Ez B, \label{Eq:ZeemanEnergy}
\end{align}
where $g_\text{eff} \equiv g_0-g_1$ plays the role of an \emph{effective} Zeeman coupling. 
% It is worth pointing out that irrespective of the sign of $g_{\text{eff}}$, the critical field $B_{\ast}$, beyond which the Landau levels crossing is avoided, is only mildly affected. A more detailed discussion on this issue is provided in Appendix~\ref{Appendix:Zeeman}.    

Notice that neither the orbital nor the Zeeman effect can lift the Ising-type symmetry between $\bX$ and $\bY$ valleys. However, the application of an external uniaxial strain field on the surface can break this symmetry. The strain field can be applied by placing a piezoelectric material on the surface with a voltage bias, and reasonable values of $\dst$ can be estimated to be about few meV~\cite{Shayegan:2006ey}. For concreteness, we assume that strain is applied along the $\bG$-$\bY$ line, so that the surface gets elongated (contracted) along the $\bG$-$\bY$ ($\bG$-$\bX$) direction. Such perturbation lowers (raises) the energy of carriers in the $\bX$ ($\bY$) valley~\cite{Shayegan:2006ey}, and consequently the degeneracy between the $\bX$ and $\bY$ valleys is lifted, which can be accounted for by setting 
\begin{align}
	\delta H_{\bX}^{\text{str}} = -\dst, \quad \delta H_{\bY}^{\text{str}} = \dst, \label{Eq:strain}
\end{align}
in Eq.~\eqref{Eq:H0}. Note that in principle the energy shift between $\bX$ and $\bY$ valleys can be different in magnitude. { Nevertheless, due to the cubic symmetry of the system, which translates into a four-fold rotational symmetry on the (001) surface, the compressibility tensor is expected to by symmetric under $x \leftrightarrow y$, and consequently $\delta H_{\bX}^{\text{str}}=\delta H_{\bY}^{\text{str}}$. }However, because our focus is on the ground state of the system, such a simplified assumption is sufficient. We will comment on this issue again in Sec.~\ref{Section:Strain}.
Aside from developing an energy offset between $\bX$ and $\bY$ valleys, the uniaxial strain can also give rise to $\bm{k}$-dependent perturbations to the effective Hamiltonian in Eq.~\eqref{Eq:H0}, which to lowest order in $\bm{k}$, only modifies the anisotropy of the three surface Dirac cones according to   
\begin{align}
	\delta_{\bG} = \delta_c, \: \: \delta_{\bX} = \delta - \delta_c, \:\: \delta_{\bY} = \delta + \delta_c.  \label{Eq:delta_c}
\end{align}
Such a change in anisotropy does not shift the energy of the zeroth Landau level, but changes the wave-functions and consequently can lead to interesting features in the phase diagrams of interacting surface states. 

Notice, in the presence of a uniaxial strain that removes the degeneracy between $\bX$ and $\bY$ valleys, quantum Hall states can be observed for $|\nu|=\frac{1}{2},\frac{3}{2},\frac{5}{2},\frac{7}{2},\cdots$.

\section{Electron-electron interaction in lowest Landau level\label{Section:Interaction}}

\begin{figure}[!]
\includegraphics[scale=0.8]{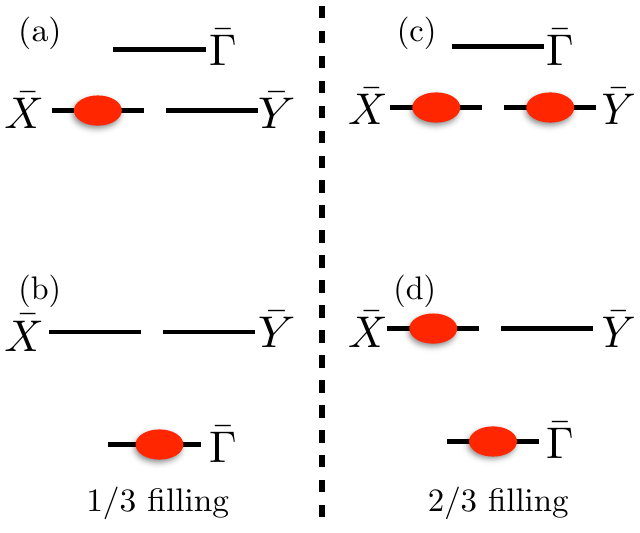}
\caption{Illustration of particle-hole symmetry in the $N=0$ Landau level subspace. In (a) and (c) the $\bG$ valley Landau level has a higher energy, while in (b) and (d) the situation is reversed. \label{Fig:PH}}
\end{figure}

We now focus on the subspace of three zeroth Landau levels ($N=0$) on the (001) surface, and study the effects of electron-electron interaction in this subspace. Since the effect of interactions in the quantum Hall regime is particularly prominent at integer Landau level fillings~\cite{Yang2006:QHFM}, we only consider the scenario when one of the three zeroth Landau levels is completely occupied. 
{Moreover, due to the particle-hole symmetry of a filled Landau level~\cite{Fertig1994,Yacoby2013}, our discussion is equally applicable at $2/3$ filling in this subspace. Specifically, the set of broken-symmetry phases at $2/3$ filling can be one-to-one mapped onto those at $1/3$ filling. 
For example, as shown in Fig.~\ref{Fig:PH} the situations in (a) and (b) are related to (d) and (c) by particle-hole symmetry, respectively. 
Thus we will not elaborate on the physics at $2/3$ filling. }

The central aspects of electron-electron interaction in the quantum Hall regime can be captured within a simplified framework, if one is allowed to neglect Landau level mixing. Such an approximation can be justified in many two-dimensional electronic systems, such as GaAs quantum wells~\cite{MacDonald:1985QHF, MacDonald:1991kw} and possibly in graphene~\cite{Bitan:2014Graphene-1,Goerbig:2011bh}. We believe that this approximation can also be justified for the surface states of THBs, for the following reason. In the quantum Hall regime the ratio of Coulomb energy to Landau level separation is equal to $e^2/\kappa \hbar v$~\cite{Goerbig:2011bh} $ \sim 14.6/\kappa$, for a given Fermi velocity of the surface Dirac fermions $v \sim 10^5$\,m/s (extracted from quantum oscillation measurements~\cite{Li:2014hk}), where $\kappa$ is the dielectric constant. Therefore, the regime of small Landau level mixing is accessible for experimentally available magnetic fields. 
%Furthermore, while the Landau level mixing plays an important role in distinguishing various broken symmetry phases in the quantum Hall regime~\cite{Bitan:2014Graphene-1}, our focus here is mainly on the distribution of electronic density among the three valleys, for which it can be safely neglected.

In the absence of Landau level mixing, the exact ground state at integer filling factors is known to be comprised of a set of fully occupied Landau level orbitals that share a common spinor describing the internal degrees of freedom such as spin, layer, or valley~\cite{MacDonald:1985QHF, MacDonald:1991kw, Sondhi:1993QHF, Moon:1995ck,Zheng1997:QHF,DasSarma1997:QHF,DasSarma1998:QHF,MacDonald:1999QH,Nomura2006:QHFM,Yang2006:QHFM,Alicea2006:QHF}. Some well known systems that can support such broken-symmetry states, often referred to as \emph{quantum Hall ferromagnets}, are GaAs and AlAs quantum wells~\cite{Spielman:2000fs, Shayegan:2006ey}, single and multilayer graphene~\cite{Feldman:2009he, Phillip:2010BLG, Zhang:2012fy, Lee:2013dd, Bitan:2014Graphene-2}, surfaces of silicon~\cite{Eng:2007gz, Kott:2014ix} and SnTe~\cite{Xiao:2015TCI-QH}, and bulk bismuth~\cite{Ong:2008Bismuth, Zhu:2012ib}. 
Here we demonstrate that in the quantum Hall regime, the ground state of surface Dirac fermions in THBs is also quantum Hall ferromagnets with various spontaneous valley orderings, which, while sharing some similarity with other 2D systems, has its own peculiar and interesting features.

{In what follows, we will completely neglect the Landau level mixing due to any ordering with the zeroth LL. Note that LL mixing plays an important role in multi-component systems such as few-layer graphene, where due to the valley and spin degeneracy of the zeroth LL, a degenerate set of broken-symmetry phases can develop within this manifold. Thus, one needs to account for LL mixing to unambiguously pin the nature of the ground states. In fact, such a mechanism turns out to be quite important and decisive within the zeroth LL in few-layer graphene. 
However, the surface of strong TI only hosts non-degenerate zeroth LLs near each valley. Therefore, electronic density can only display either valley-polarized or valley-coherence distributions for a uniform state. Incorporation of LL mixing effect can thus only alter the outcomes quantitatively, without affecting the qualitative features predicted in our work.}

\subsection{Hartree-Fock Hamiltonian}

Our analysis is based on the Hartree-Fock approximation, which happens to be exact at integer Landau level filling factors for large enough applied magnetic fields~\cite{MacDonald:1991kw,Yang2006:QHFM}. We assume that the system has a uniform ground state, so that the Hartree contribution cancels with the positive background. As a result, the only contribution comes from the Fock (i.e. the exchange) term, yielding the mean-field Hamiltonian~\cite{MacDonald:1991kw},
\begin{align}
	H_{\text{MF}} = &-\dfrac{N_{\phi}}{S}\sum_{\bq,\lambda_{i}} V(\bq)F_{\lambda_1\lambda_4}(\bq)F_{\lambda_2\lambda_3}(-\bq){\rho}_{\lambda_{1}\lambda_{3}} \average{{\rho}_{\lambda_{2}\lambda_{4}}} \notag\\ & \times e^{iq_xq_y\ell^2 (1-w_{\lambda_1\lambda_4}-w_{\lambda_2\lambda_3})}, \label{Eq:FockHamiltonian}
\end{align}
where $S$ is the surface area of $(001)$ plane, $N_{\phi} = S/2\pi\ell^2$ is the orbital degeneracy of each Landau level, and $\rho_{\lambda_{1}\lambda_{4}}$are components of the \emph{density matrix}. $F_{\lambda_1\lambda_4}(\bq)$ are the \emph{form factors}~\cite{Yang2006:QHFM}, which, when projected onto the subspace of three $N=0$ Landau levels, is given by  
\begin{align}
	F_{\lambda_j \lambda_k}(\bq) =\sqrt{\dfrac{2\alpha_{\lambda_j}\alpha_{\lambda_k}}{\alpha_{\lambda_j}^2+\alpha_{\lambda_k}^2}} e^{-\bar{q}^2_{\lambda_j \lambda_k}\ell^2/2}, 
\end{align}
where 
\begin{equation}
\bar{q}^2_{\lambda_j \lambda_k} = \frac{{\alpha_{\lambda_j}^2\alpha_{\lambda_k}^2q_x^2+q_y^2}}{{\alpha_{\lambda_j}^2+\alpha_{\lambda_k}^2}}.
\end{equation} 
In addition, anisotropy of the Dirac cones gives rise to a new contribution in $H_{\text{MF}}$,
\begin{equation}
w_{\lambda_j\lambda_k}=\frac{\alpha_{\lambda_j}^2}{{\alpha_{\lambda_k}^2+\alpha_{\lambda_k}^2}}.
\end{equation}
In the presence of electron-electron interaction, the ground states within this subspace can then be obtained from the mean-field Hamiltonian in Eq.~\eqref{Eq:FockHamiltonian}. In what follows we obtain explicit forms of $H_{\text{MF}}$ for the following two cases: (i) intravalley interaction involving small momentum transfer, and (ii) intervalley interactions for which the (large) momentum transfer is proportional to the separation between the valleys in the momentum space.

\subsubsection{Intravalley scattering}

 Among various interaction channels, the leading contribution comes from the \emph{intravalley} interaction, for which scattering between two electrons occurs within the same valley~\cite{Goerbig:2011bh}. Contribution from such a process is captured by setting $\lambda_1=\lambda_4=\lambda$ and $\lambda_2=\lambda_3 = \sigma$ in Eq.~\eqref{Eq:FockHamiltonian}. The mean-field Hamiltonian then takes the form  
\begin{align}
	H_{\text{MF}}^{\text{intra}} = -N_{\phi}\dfrac{e^2}{\kappa\ell} \rho_{\lambda\sigma}\average{\rho_{\sigma\lambda}} X_{\lambda\sigma}, \label{Eq:FockHamiltonian-LR}
\end{align}
where the superscript ``intra'' stands for intravalley contributions. The elements of exchange integral matrix  are
\begin{align}
	X_{\lambda\sigma} = \dfrac{\kappa\ell}{e^2S}\sum_{\bq} \dfrac{2\pi e^2}{\kappa q} F_{\lambda\lambda}(\bq)F_{\sigma\sigma}(-\bq),  \label{Eq:CoefficientX}
\end{align}
after replacing $V(\bq)$ in Eq.~\eqref{Eq:FockHamiltonian} by the 2D Coulomb potential ($V(\bq) \sim e^2/|\bq|$). Although actual values of exchange integrals depend on the anisotropy factor $\delta$ of the Fermi surfaces (see Table~\ref{Table:ExchangeIntegral}), some general features of $X_{\lambda\sigma}$ can be appreciated based on symmetries: (i) The four-fold rotational symmetry between $\bX$ and $\bY$ valleys guarantees that $X_{22} = X_{33}$ and $X_{12} = X_{13}$. Therefore, $X_{\lambda\sigma}$ has only \emph{four} independent elements, namely $X_{11}$, $X_{22}$, $X_{12}$ and $X_{23}$. (ii) The Fermi surface ellipticity reduces the strength of the exchange integral~\cite{Xiao:2015TCI-QH} for long-range Coulomb interaction. Hence, the two diagonal elements satisfy $X_{22}<X_{11}$ in general, and the difference $X_{11}-X_{22}$ increases with increasing $\delta$, as can be seen from Table~\ref{Table:ExchangeIntegral}.

\begin{table}[!]
\begin{tabular}{|l|c|c|c|c|}
\hline\hline
             & $X_{11}$ & $X_{22}$ & $X_{12}$ & $X_{23}$ \\ \hline
$\delta=0$   & 0.8862   & 0.8862   & 0.8862   & 0.8862   \\ \hline
$\delta=0.1$ & 0.8862   & 0.8840   & 0.8834   & 0.8774   \\ \hline
$\delta=0.2$ & 0.8862   & 0.8772   & 0.8750   & 0.8515   \\ \hline
$\delta=0.3$ & 0.8862   & 0.8654   & 0.8604   & 0.8098   \\
\hline\hline
\end{tabular}
\caption{Typical values for the exchange integral in Eq.~\eqref{Eq:CoefficientX}. \label{Table:ExchangeIntegral}}
\end{table}

\subsubsection{ Intervalley interaction or backscattering }

In THBs, orbitals with relatively high angular momentum (such as $p$, $d$, $f$ orbitals) participate in the bulk band-inversion. Consequently the surface states are also composed of orbitals in which Hubbard-type short-range interactions can be substantially strong. The presence of such strong residual short-range interactions on the surface states calls for the inclusion of large-momentum backscattering in the theory, which enters the Hamiltonian in the form of \emph{intervalley} interactions~\cite{Goerbig:2011bh}. This contribution can be analyzed by taking $\lambda_1=\lambda_3 = \lambda$ and $\lambda_2=\lambda_4 = \sigma$ in Eq.~\eqref{Eq:FockHamiltonian}, for which the mean-field Hamiltonian takes a rather compact form 
\begin{align}
	H_{\text{MF}}^{\text{BS}} = -\dfrac{N_{\phi}e^2}{\kappa\ell} \sum_{\lambda\neq\sigma}\rho_{\lambda\lambda}\average{\rho_{\sigma\sigma}} Y_{\lambda\sigma}, \label{Eq:FockHamiltonian-SR}
\end{align}
where the superscript ``BS'' denotes contributions from backscattering. The matrix elements $Y_{\lambda\sigma}$ are given by 
\begin{align}
	Y_{\lambda\sigma} &= \dfrac{\ell}{e^2S}U_{\lambda\sigma}\sum_{\bq} F_{\lambda\sigma}(\bq)F_{\sigma\lambda}(-\bq) = \dfrac{U_{\lambda\sigma}}{2\pi e^2\ell}, \label{Eq:CoefficientY}
\end{align}
where $U_{\lambda\sigma}$ represents the strength of intervalley scattering processes between two valleys $\lambda$ and $\sigma$. Although it is challenging to get an accurate estimation of $U_{\lambda\sigma}$, it may not be too unrealistic to take $U_{\lambda \sigma} \sim {2\pi e^2}/{K_{\lambda\sigma}}$, where $K_{\lambda\sigma}$ is the momentum-space separation between valleys $\lambda$ and $\sigma$. The underlying rotational symmetry between $\bX$ and $\bY$ valleys guarantees that $Y_{12}=Y_{13}$, similar to the situation for $X_{\lambda \sigma}$. Consequently, the Hamiltonian in Eq.~\eqref{Eq:CoefficientY} is defined in terms of only two independent parameters, $Y_{12}$ and $Y_{23}$. We also note while $Y_{\lambda\sigma}$ scales as $\sqrt{\Bperp}$, $X_{\lambda\sigma}$ is independent of the applied magnetic field. This difference will play an important role in determining the ground states of the system.

\subsection{Mean-field energy functional}

We can now proceed with Eqs.~\eqref{Eq:FockHamiltonian-LR} and~\eqref{Eq:FockHamiltonian-SR} and arrive at the mean-field energy functional within the subspace of three zeroth Landau levels. For convenience, we introduce a three-component spinor $\Psi = [z_1, z_2, z_3]$ to describe possible ground states, where complex variables $z_\lambda = r_\lambda e^{i\theta_\lambda}$ for $\lambda=1,2,3$, respectively describe the electron occupation in the $\bG$, $\bX$, and $\bY$ valleys. The mean-field energy functional then becomes 
\begin{eqnarray} 
	E_{\text{MF}} &= N_{\phi}\biggl(\Delta_0 r_1^2 - \dfrac{e^2}{\kappa\ell} \biggl[ \displaystyle\sum^{3}_{j=1} X^2_{jj} r^4_j \notag\\ 
	&+ \displaystyle\sum_{j,k=1}^{3}{'} \left( X_{jk}+Y_{jk}\right) r^2_j r^2_k \biggr] \biggr), \label{Eq:EnergyFunctional}
\end{eqnarray}
which includes contributions from both the single-particle and the Fock terms. The notation $\sum'$ indicates that terms with $j=k$ are excluded from the summation. Note that phases ($\theta_\lambda$) of all complex variables $z_\lambda$ drop out from the mean-field energy functional $E_{\text{MF}}$, indicating that the electron number in each valley is separately conserved, and thus the mean-field Hamiltonian exhibits a U(1)$\otimes$ U(1) $\otimes$ U(1) symmetry.

To capture various possible phases in this system, we introduce a new set of variables 
\begin{align}
	r_1 = \sqrt{1-R^2}, \; r_2 = R\cos\phi, \; r_3 = R\sin\phi, 
\end{align}
with $0\leq R\leq1$ and $0\leq \phi \leq \pi/2$. With such a parametrization, $R=0$ represents a state polarized in the $\bG$ valley. On the other hand, polarization of electronic density in the $\bX$ and the $\bY$ valley correspond to $R=1$, with $\phi=0$ and $\pi/2$, respectively. A valley-coherent distribution of electronic density between $\bX$ and $\bY$ is represented by $R=1$ with $0<\phi<\pi/2$. Finally, $0<R<1$ indicates a coherent superposition between $\bG$ and $\bX$ or $\bY$ valleys. Therefore, different phases can be identified quite effectively through this parametrization. In terms of these new parameters ($R$ and $\phi$), the mean-field energy functional in Eq.~\eqref{Eq:EnergyFunctional} assumes a compact form  
\begin{align}
	 E_{\text{MF}}(R, \phi) = \frac{1}{4\kappa} \left( a \; R^4+ b \; R^2+c \right), \label{Eq:Minimization}
\end{align}
where 
\allowdisplaybreaks[4]
\begin{align}
	a &= \sqrt{\Bperp}\Ec \bigl[(X_{23}+Y_{23}-X_{22})\cos(4\phi)\notag\\
	& \quad +(8X_{12}+8Y_{12}-4X_{11}-3X_{22}-X_{23}-Y_{23})\bigr], \notag \\
	b &= 8\sqrt{\Bperp}\Ec(X_{11}-X_{12}-Y_{12})-4\kappa \Delta_0, \notag \\
	c &= 4\kappa \Delta_0-4\sqrt{\Bperp}\Ec X_{11}. \notag
\end{align}
$\Ec \!=\! e^2/\SI{25.6}{nm} \!\simeq\! \SI{56}{meV}$ is the scale of the Coulomb interaction. The ground state of the system is then obtained by minimizing the energy functional in Eq.~\eqref{Eq:Minimization}.

\section{Spontaneous valley ordering in zeroth Landau level\label{Section:GroundStates}}

We proceed with the mean-field energy functional from Eq.~\eqref{Eq:Minimization} and search for possible ground states resulting from electron-electron interaction. We also explore the influence of Zeeman coupling and strain field. However, it is worth incorporating these ingredients one by one, as they influence the phase diagram differently. 

\subsection{Pristine surface without Zeeman or strain field}

\begin{figure}[!]
\includegraphics[scale = 1.1]{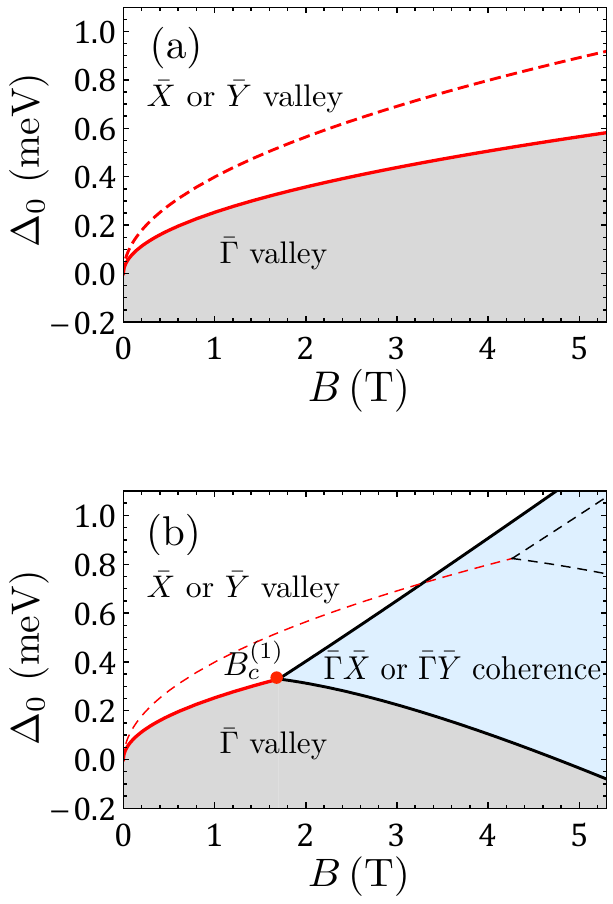}
\caption{Phase diagram within the zeroth Landau level of interacting surface states, residing on the (001) surface of THBs in the presence of (a) intravalley scattering and (b) both intravalley and backscattering interactions. We here set $\delta=0.2$, $\kappa=2$, and the backscattering interaction $U_{\lambda\sigma}=2\pi e^2/K_{\lambda\sigma}$. Red (black) lines represent first-order (continuous) phase transition. The red dot in (b) marks the critical field $B_{c}^{(1)}$ defined in Eq.~\eqref{Eq:FirstBfield}. The dashed lines are obtained for $\delta=0.25$, while keeping the rest of the parameters in the theory unaltered. \label{Fig:PhaseCoulomb}}
\end{figure}

Let us begin the discussion by turning off Zeeman coupling and external strain field, and solely focus on the effects of electron-electron interaction. This exercise will provide valuable insights into the overall structure of the phase diagram of interacting surface states, as shown in Fig.~\ref{Fig:PhaseCoulomb}. 
We first keep only the contributions from intravalley interaction, which amounts to setting all $Y_{\lambda\sigma}$ to zero in Eq.~\eqref{Eq:Minimization}. In such a limit, electrons maximize the exchange energy gain by keeping only one of three valleys fully occupied. Therefore, the ground state must always be valley polarized. In particular, in the absence of an energy offset among the valleys (i.e., $\Delta_0=0$), electrons must occupy the $\bG$ valley, which continues as the preferred ground state for arbitrary $\Delta_0<0$, as shown by the shaded region in Fig.~\ref{Fig:PhaseCoulomb}(a) for $\delta=0.2$. Such an outcome can be anticipated intuitively, since a circular Fermi surface acquires maximal exchange energy from the long range tail of the Coulomb interaction (notice $X_{11}>X_{12}$ in Table~\ref{Table:ExchangeIntegral} for any $\delta \neq 0$).

It is useful to contrast the current situation with 2D systems like graphene~\cite{Yang2006:QHFM} and monolayer MoS$_{2}$~\cite{Li:2013tx}, where all Fermi pockets are isotropic. In those systems intravalley Coulomb interaction by itself cannot determine the ground state uniquely. This outcome can be appreciated from the energy functional Eq.~\eqref{Eq:EnergyFunctional} if we set both $\delta$ and $Y_{\lambda\sigma}$ to zero. Under that circumstance, all entries from the intravalley exchange integral are equal [see Table~\ref{Table:ExchangeIntegral}], and the contributions to the energy functional Eq.~\eqref{Eq:EnergyFunctional} become constant, independent of the ground state configuration. Indeed, the actual ground state in graphene is likely to be determined by interaction effects at the lattice scale~\cite{Herbut2007:GrapheneIQHE,Herbut2007:GrapheneSO(3),Kharitonov2012:Graphene-QHE,Wu2014:Graphene-QHE, Bitan:2014Graphene-1,Bitan:2014Graphene-2}.

When the noninteracting zeroth Landau level near $\bG$ valley is placed at a higher energy than those residing near $\bX$ and $\bY$ valleys (i.e., $\Delta_0>0$), the single-particle energy offset ($\Delta_0$) competes with the exchange energy, and there exists a first-order phase transition from a $\bX$ (or $\bY$)-valley polarized state to a $\bG$-polarized state as the field strength is increased. Notice that $\bX$ and $\bY$ valleys are always degenerate and therefore the $\bar{X}$- (or $\bar{Y}$-) valley polarized state spontaneously breaks an Ising symmetry. The finite-temperature phase transition into this phase belongs to the two-dimensional Ising universality class, and the ground state represents a \emph{quantum Hall Ising nematic order}, as the four-fold rotation symmetry on the surface is spontaneously broken.  

The boundary of the aforementioned first-order transition is determined by 
\begin{align}
	\Delta_0 = \Ec(X_{11}-X_{22})\sqrt{\Bperp}/\kappa, \label{Eq:Phaseboundary-I}
\end{align}
in $B$-$\Delta_0$ plane and shown by the red lines in Fig.~\ref{Fig:PhaseCoulomb}(a) for $\delta=0.2$ (solid line) and $\delta=0.25$ (dashed lineb). The \emph{square-root} field dependence of this phase boundary stems from the $\sqrt{B}$ scaling of the exchange energy from intravalley interactions. With increasing ellipticity of $\bX$ and $\bY$ valleys ($\delta$), the difference between diagonal components of the exchange integrals $(X_{11}-X_{22})$ increases and the phase boundary gets shifted upward in the $B$-$\Delta_0$ plane [compare the solid and dashed lines in Fig.~\ref{Fig:PhaseCoulomb}(a)].

With the inclusion of backscattering interaction, the possibility of realizing a valley-coherent distribution of electronic density also arises. In the presence of backscattering, the line of first-order transition ceases to exist beyond a critical field $B_c^{(1)}$ [shown by the \emph{red} dot~\footnote{Note that the point marked by this red dot is actually a \emph{tricritical} point in the phase diagram, at which a line of first-order transition terminates and two lines of continuous phase transition emanate.} in Fig.~\ref{Fig:PhaseCoulomb}(b)], and bifurcates into two lines of continuous phase transitions [shown by the black lines in Fig.~\ref{Fig:PhaseCoulomb}(b)]. The critical field is given by  
\begin{align}
	 B_c^{(1)} = \frac{\hbar \pi^2 e^3}{U^{2}_{12}}  \; (X_{11}+X_{22}-2X_{12})^2, \label{Eq:FirstBfield}
\end{align}
yielding $B_c^{(1)}\sim\SI{1.7}{T}$ and $\SI{4.3}{T}$, respectively for $\delta=0.2$ and $0.25$, and $U_{12} = 2\pi e^2/K_{12}$. The quantity in the parentheses in Eq.~\eqref{Eq:FirstBfield} increases with $\delta$, and hence the critical field $B_c^{(1)}$ also increases as the pockets near $\bX$ and $\bY$ points become more elliptic.

The lines of continuous phase transitions in the $B$-$\Delta_0$ plane are obtained from  
\begin{align}
	 \Delta_0 =& \frac{2\sqrt{B} \Ec}{\kappa} \left( X_{11}-X_{12}-\dfrac{U_{12}}{2\pi e^2\ell} \right), \label{Eq:PB-GGX} \\ 
	 \Delta_0 =& \frac{2\sqrt{B} \Ec}{\kappa} \left( X_{12}-X_{22}+\dfrac{U_{12}}{2\pi e^2\ell} \right). \label{Eq:PB-GX-X} 
\end{align}
Equation~(\ref{Eq:PB-GGX}) defines the phase boundary between a pure $\bG$-polarized state and a state with coherent superpositions between $\bG$ and $\bX/\bY$-valley. In contrast, the phase boundary between $\bG$-$\bX/\bY$ coherent phase and pure $\bX$/$\bY$ polarized state is determined from Eq.~(\ref{Eq:PB-GX-X}). Notice that unlike the transitions between valley-polarized phases, these phase transitions are deemed continuous, since the ground-state wave function does not display any abrupt changes across these boundaries.

For the valley-coherent phase, shown in Fig.~\ref{Fig:PhaseCoulomb}, aside from the $Z_2$-Ising symmetry between $\bX$ and $\bY$ valleys, the continuous U(1) symmetry associated with the conservation of electronic density at each valley is also spontaneously broken. As a result, the collective excitations in such valley-coherent phase are gapless Goldstone modes. 

In principle, it is conceivable to realize a ground state where the electronic density is coherently shared between $\bX$ and $\bY$ valleys. The critical field ($B_c^{(2)}$) above which such a phase may develop in the system is  
\begin{align}
  B_c^{(2)}=\frac{\hbar \pi^2 e^3}{U^2_{23}} \: (X_{22}-X_{23})^2. \label{Eq:SecondBfield}
\end{align}
However, with a reasonable estimation of intervalley interaction $U_{23}\!\sim\! 2 \pi e^2/K_{23}$, we find $B_c^{(2)}\!=\!\SI{50}{T}$ for $\delta=0.2$. This is the reason such a phase does not appear in Fig.~\ref{Fig:PhaseCoulomb}(b). Relatively high value of $B_c^{(2)}$ (in comparison to $B_c^{(1)}$) originates from two sources. First, the separation between $\bX$ and $\bY$ valleys is larger than that between $\bG$ and $\bX$ or $\bY$ valleys. Therefore, it is natural to anticipate that the intervalley interaction between $\bX$ and $\bY$ pockets, $U_{23}$, is weaker than $U_{12}$. Second, the difference among the exchange integrals appearing on the right-hand-side of Eq.~\eqref{Eq:SecondBfield} is also much larger than that of Eq.~\eqref{Eq:FirstBfield}. 

\begin{figure}[!]
\includegraphics[scale = 1.1]{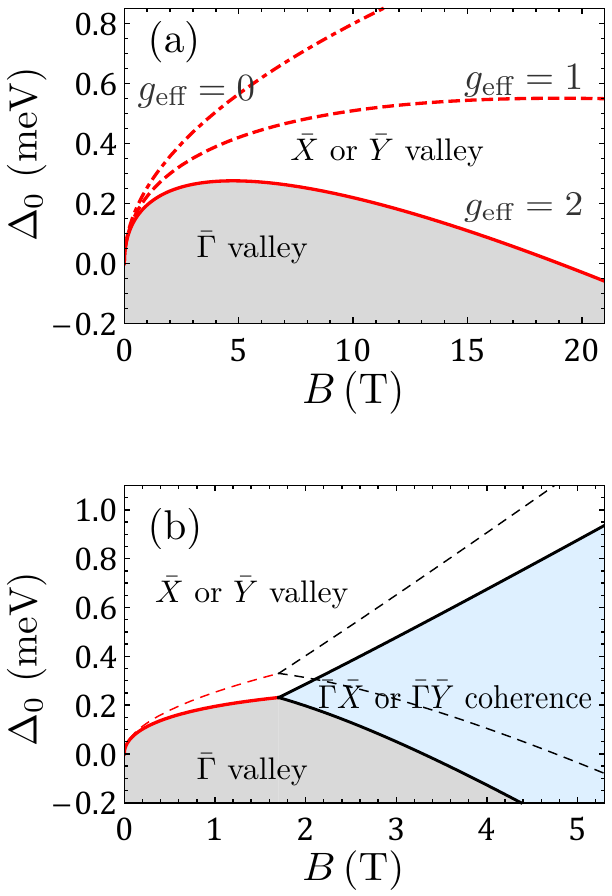}
\caption{Phase diagram with (a) only intravalley and (b) both intravalley and backscattering interactions for various values of $g_{\text{eff}}$. For $g_{\text{eff}}=0$, we recover the phase diagram without the Zeeman effects [the red line Fig.~\ref{Fig:PhaseCoulomb}(a)]. In (b) we set the $g_{\text{eff}}=2$, and the dashed lines reproduce the phase boundaries of Fig.~\ref{Fig:PhaseCoulomb}(b). We set $\delta=0.2$ and $\kappa=2$ for all plots. \label{Fig:PhaseZeeman}}
\end{figure}

\subsection{Phase diagram with Zeeman coupling}

We will now discuss the effects of Zeeman coupling on the phase diagram. We show that it does not alter the qualitative nature of the phase diagram, instead plays the role of a tuning parameter that allows us to access various regions of the phase diagram. Our findings are summarized in Fig.~\ref{Fig:PhaseZeeman}, which are obtained by replacing the energy offset $\Delta_0$ in Eq.~\eqref{Eq:EnergyFunctional} by $\Delta_0^{Z}$ (energy offset after incorporating the Zeeman coupling) from Eq.~\eqref{Eq:ZeemanEnergy}. We note that, although the Zeeman coupling is weak at small fields, it becomes important as the magnetic field gets stronger since it grows faster with the applied field than the intravalley Coulomb energy.

Indeed, the results in Fig.~\ref{Fig:PhaseZeeman} fulfill this expectation. When the effective Zeeman coupling $g_{\text{eff}}=g_{\bG}-g_{\bX/\bY}>0$, it can cause a transition out of the $\bG$-polarized state into an $\bX$- or $\bY$-polarized state for a moderately strong magnetic field, which was previously inaccessible just by tuning the field strength [see Fig.~\ref{Fig:PhaseCoulomb}(a)]. Such an effect is transparent in Fig.~\ref{Fig:PhaseZeeman}(a), where the intervalley interaction is switched off. Even when we incorporate backscattering, the qualitative structure of the phase diagram remains unchanged, as shown in  Fig.~\ref{Fig:PhaseZeeman}(b). However, the regime of $\bX$- or $\bY$-valley polarized as well as $\bG$-$\bX/\bY$ valley coherent phases increases in the phase diagram with increasing $g_{\text{eff}}$. 
Thus, by changing the Zeeman coupling, which one can tune to a certain degree by tilting the magnetic field away from the perpendicular orientation (since the Zeeman coupling depends on the total magnetic field) on (001) surface, various regions of the phase diagram and associated phase transitions can be accessed. 
In contrast, the situation for $g_{\text{eff}}<0$ is less interesting, as the phase diagram does not qualitatively differ from that obtained without the Zeeman coupling. 

\subsection{Effects of uniaxial strain field\label{Section:Strain}}

Finally, we investigate the effects of strain on the phase diagram. We restrict ourselves with a uniaxial strain applied along the $\bG$-$\bY$ direction, which breaks the rotational symmetry between $\bX$ and $\bY$ valleys. 
Hence, the application of such a uniaxial strain lifts the Ising-type symmetry between $\bX$ and $\bY$ valleys. As shown in Sec.~\ref{Section:ZeemanStrain}, such a strain field has two distinct consequences. We argue that both of them favor $\bX$ valley polarization as the ground state of the system. 
{Note that due to the underlying cubic symmetry of the system, we have assumed the strain-induced energy shifts to be equal in magnitude for $\bX$ and $\bY$ pockets, as shown in Eq.~\eqref{Eq:strain}. Even for the sake of argument, we allow the shifts in energy to be different for $\bX$ and $\bY$ valleys, the valley, pushed to the higher energy, will always be excluded from the ground state manifold, irrespective of its magnitude.}

\begin{figure}[!]
\includegraphics[scale = 1.1]{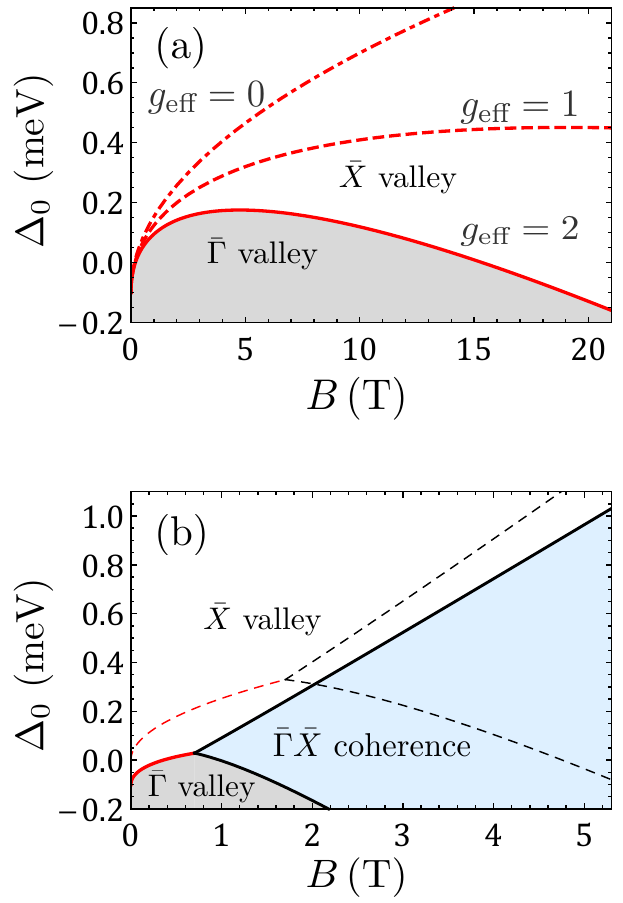}
\caption{Phase diagrams with Zeeman coupling and uniaxial strain fields for (a) pure intravalley and (b) both intravalley and backscattering interactions. Perturbations from strain fields are $\Delta_{\text{st}} = \SI{0.1}{meV}$ and $\delta_{c} = 0.02$, which places the $\bY$ valley at a higher energy. Rest of the parameters are the same as those in Fig.~\ref{Fig:PhaseZeeman}. The dashed lines in (b) reproduces the solid lines in Fig.~\ref{Fig:PhaseCoulomb}(b) for comparison. \label{Fig:PhaseStrain}}
\end{figure}

The first one arises from a single-particle effect that increases (decreases) the energy of the $\bY$ ($\bX$) valley by $\Delta_{\text{st}}$ [see Eq.~\eqref{Eq:strain}] and thus drives the $\bY$ valley out of the ground state manifold. The second one is more subtle. As shown in Eq.~\eqref{Eq:delta_c}, the applied strain field reduces (increases) the anisotropy of the Dirac cones at $\bX$ ($\bY$) valley. According to Table~\ref{Table:ExchangeIntegral}, this effect amounts to increasing (reducing) the exchange energy gain for electrons in the $\bX$ ($\bY$) valley, which further pushes the $\bY$ valley to a higher energy (relative to the $\bX$ valley). Consequently, the $\bY$ valley no longer appears in the ground state phase diagram when a uniaxial strain is applied along $\bG$-$\bY$ on the surface, as shown in Fig.~\ref{Fig:PhaseStrain}. In addition, in comparison with Fig.~\ref{Fig:PhaseZeeman} we find that the strain field gives rise to a larger phase space for the $\bX$-valley polarized and $\bG$-$\bX$ coherent phases. 
Such an outcome stems from the fact that uniaxial strain (applied along $\bG$-$\bX$ or $\bG$-$\bY$) also turns the $\bG$ valley into an elliptic one, thus reducing its energy gain from electron-electron interaction [see Eq.~\eqref{Eq:delta_c}] (also note that the ellipticity of $\bX$ valley is reduced at the same time). Therefore, an applied strain on the surface of THBs serves as a useful tuning parameter to explore various broken symmetry phases. 

\subsection{{Pure short-range interactions}}
Before concluding this section a discussion on screening of long-range Coulomb interaction seems appropriate. 
Let us replace the long-range Coulomb potential in Eq.~\eqref{Eq:FockHamiltonian-LR} by a short-ranged one $2\pi e^2/k_0$, where the constant $k_0$ characterizes the strength of the short-range interaction. 
With such an interaction potential, all diagonal elements of the exchange integral $X_{\lambda\lambda}$ are identical, with $X_{11}=X_{22}=X_{33}=1/(2k_0\ell)$. The off-diagonal elements are still smaller than the diagonal ones, given by $X_{12}$\,$=$\,$X_{13}$\,$=$\,$ \sqrt{6}/(5k_0\ell)$ and $X_{23} $\,$=$\,$ 6/(13k_0\ell)$. Therefore, when the long-range Coulomb interaction is replaced by a short-ranged one, circular Fermi pockets lose the energetic advantage over elliptic ones. Consequently, when the energy off-set $\Delta_0$ is zero, the three valleys always remain degenerate, and the electronic density spontaneously chooses one of them to condense into at low temperatures. The finite-temperature phase transition into such a broken-symmetry ground state is described by the \emph{three-state Potts model}. Near the phase transition the correlation length exponent is $\nu=5/6$ and the order parameter anomalous dimension is $\eta=4/15$~\cite{Pottsmodel}. In contrast, for arbitrary $\Delta_0<0$ and $\Delta_0>0$ the ground state is respectively $\bG$-valley and $\bX/\bY$-valley polarized. Thus, with pure short range interaction and for $\Delta_0>0$, the ground state corresponds to a quantum Hall Ising-nematic order. 

% \subsection{The $(111)$ surface}
% \edit{Finally we focus on yet another high symmetry surface of cubic TKIs, the (111) surface. The BZ of (111) surface is shown in Fig.??, which accommodates three degenerate copies of massless Dirac cones (no energy offset among the Dirac points). The surface LLs display an exact three-fold degeneracy, as shown in Fig.??. Consequently, as one tunes the surface chemical potential for fixed magnetic field, or vice-versa, a series of Hall plateaus is expected to appear at fillings $\nu=3 \left( \frac{1}{2}+N \right)$, for $N=0,1,2, \cdots$ (augmented by a factor 2 due to the presence of top and bottom surfaces). Therefore, from the quantization of Hall plateaus one can pin the actual orientation of the surface. 

% When electron-electron interactions are taken into account on (111) surface, and electronic density is such that only one of the three degenerate zeroth LL can be occupied there are three possible broken symmetry phases: (1) X1, X2 or X3- valley polarized state, (2) Valley coherent X1-X2, X2-X3, X3-X1 phases, (3) X1-X2-X3 valley coherent phase. Notice three possible valley polarized and valley coherent [in class (2)] phases are three-fold degenerate. Therefore, a finite temperature phase transition into either of these two phases belongs to the universality class of the three-state Potts model. On the other hand, X1-X2-X3 valley coherent phase only lacks the rotational symmetry in general, and corresponds to a nematic order. }

\section{Discussion and Conclusion\label{Section:Discussions}}

To summarize, in this work we have studied the interplay of external magnetic fields (as well as strain) and electron-electron interaction on the two-dimensional surface states of a family of heavy fermion hexaboride compounds which are predicted to be topological insulators~\cite{neupane:2014}. Systems that may support such topologically nontrivial ground states at low temperatures are SmB$_6$, YbB$_6$, PuB$_6$. These materials have a bulk insulating gap, share a common cubic symmetry, and the band inversion (requisite criteria for an insulator to be topological) occurs around three high-symmetry $X$ points in the bulk Brillouin zone. Consequently, the surface states are composed of three copies of massless Dirac fermions, referred to as the \emph{valleys}. For concreteness, we here focused only on the high symmetry (001) surface of THBs.

Subject to strong magnetic fields, these surface states support three copies of Dirac Landau levels. We show that depending on the relative energies of the Dirac points and their effective Fermi velocities, a variety of integer quantum Hall states can be realized (including contributions from both top and the bottom surfaces that simultaneously undergo Landau level quantization). In particular, due to the two-fold degeneracy between $\bX$ and $\bY$ valleys, we expect an absence of quantum Hall states at some filling factors. However, the application of a uniaxial strain (along $\bG$-$\bX$ or $\bG$-$\bY$) removes such a degeneracy, and quantum Hall states at all (integer) filling factors should be accessible. Robust quantum Hall states can be observed from the quantization of Hall conductivity~\cite{Xu:2014NatPhys} or the time-dependence of charge transfer from the inner to the outer edge in a Corbino-shaped system~\cite{dogopolov}.

The electron-electron interaction is shown to support a rich variety of broken symmetry phases within the zeroth Landau level. For example, if we consider only the long ranged tail of the Coulomb interaction [$V(\mathbf{q}) \sim e^2/|\mathbf{q}|$], the ground state is always a valley polarized quantum Hall ferromagnet. Long-range Coulomb interaction prefers to keep the $\bG$ valley fully occupied due to the circular Landau orbitals in its vicinity. 
Due to a four-fold rotational symmetry, $\bX$- and $\bY$-valley polarized states are energetically degenerate. Thus, at low temperatures when $\bX/\bY$ valley is filled, electronic density spontaneously breaks a valley-Ising symmetry and the ground state should represents a quantum Hall Ising nematic order. In contrast the short-range intervalley interactions can give rise to valley-coherent states, where electronic density is distributed among two valleys. While the transition between the two valley-polarized states is always first order in nature, that between a valley-coherent and valley-polarized state is continuous.

We also address the effects of Zeeman coupling and application of a uniaxial strain on the phase diagram of interacting surface states. Neither of these two external fields changes the qualitative structure of the phase diagram but modifies the relative phase space available for distinct symmetry breaking phases (valley coherent and polarized). Therefore, both Zeeman coupling (which can be tuned efficiently by tilting the magnetic field) and the uniaxial strain (introduced by placing a piezoelectric material on the surface) serve as useful tuning parameters to access various ordered phases as well as phase transitions in experiments. However, uniaxial strain breaks the Ising-type symmetry among $\bX$ and $\bY$ valleys and acts as an external field to the order parameter (electronic density in one valley). Therefore, a uniaxial strain prefers one of these two valleys as the ground state. 

% \edit{Although in this work we focused on the subspace of the zeroth LL, our analysis can be generalized to the situation when the chemical potential is tuned to higher LLs. Irrespective of the LL index there are only two possibilities: (a) valley polarized, and (b) valley coherent phases. However, for a given magnetic field, LL from different branches with different LL indices can, in principle, be energetically almost degenerate. Consequently, these two orderings can be realized in either (a) LL polarized, or (b) coherent LL superposition channels. 
% One can also envision various spatially  inhomogeneous Hartree-Fock trail ground states in the higher LLs, such as charge-density-wave states and stripe phases. However, we will not discuss these possibilities in any detail here. 
% }

Detection of the precise ground state in multi-valley systems can be challenging. We believe that a measurement of the \emph{longitudinal} resistivity ($R_{jj}$), where $j=x$ or $y$ ($z$ is the direction perpendicular to the surface), at finite temperatures and its scaling with the magnetic field can provide distinguishing signatures for various broken-symmetry phases. For example, $R_{jj}$ is expected to be isotropic in the $\bG$-polarized phase, and thus $\Delta R=|R_{xx}-R_{yy}| =0$. 
In contrast, an $\bX$ or $\bY$-polarized state should display an anisotropy between $R_{xx}$ and $R_{yy}$, but $\Delta R$ in this phase is independent of the magnetic field strength. The valley-coherent phase (either $\bG$-$\bX$ or $\bG$-$\bY$) also gives rise to anisotropic longitudinal resistivity. 
However, the relative occupation of electronic density between the two valleys ($\bG$ and $\bX$ or $\bY$) changes with the magnetic field. Therefore, $\Delta R$ is expected to change with the strength of the magnetic field in a valley-coherent phase. Hence, from the scaling of $\Delta R$ with magnetic field, it should be possible to figure out the exact pattern of symmetry breaking in the ground state.   
Similar proposals have been discussed to detect nematic valley ordering in other quantum Hall systems~\cite{Abanin2010:QHF,Kumar2013}. 

We now discuss an interesting internal structure of the zeroth Landau level that is solely tied with the topology of the surface Dirac cones. While arriving at Eq.~(\ref{Eq:H0}), we have assumed that all three Dirac cones have identical spin-momentum locking. Consequently, three copies of zeroth Landau levels have identical spin polarizations. However, it is conceivable that spin-momentum locking of the Dirac cones near the $\bG$ valley is opposite to that near $\bX$ and $\bY$ valleys. This scenario is protected by the underlying nontrivial $Z_2$ topological invariant in the bulk and can be accomplished by tuning the relative strength of nearest-neighbor and next-nearest-neighbor hybridization among the opposite parity bands~\cite{Roy:2014vv, sigrist}. When spin-momentum locking at $\bG$ and $\bX/\bY$ points is opposite, the spin-polarization of the zeroth Landau levels near these valleys is also opposite. Consequently, the $\bG$-polarized and $\bX/\bY$-polarized states are characterized by opposite spin-polarization. An even more interesting situation arises in the valley-coherent phase, where the electronic density is shared between two zeroth Landau levels from different valleys. Since the zeroth Landau level at the two valleys now carries opposite spin angular momentum, the valley-coherent state corresponds to a \emph{ferrimagnet}, with finite uniform and staggered magnetization. In this situation, effects of uniaxial strain remain unchanged, but significant modifications can arise from the Zeeman coupling. Since the $g$-factors within the zeroth Landau level at $\bG$-valley ($g_0$) and $\bX/\bY$-valley ($g_1$) are of opposite signs, the system can experience a large effective $g$-factor ($g_{\text{eff}} \equiv g_0-g_1$) even for small $g_0$ and $g_1$. Therefore, various regions of the phase diagrams in Figs.~\ref{Fig:PhaseZeeman} and \ref{Fig:PhaseStrain} can now be accessible even for weaker magnetic fields. Even when the zeroth Landau levels near different valleys carry opposite spin polarization, measurements of $\Delta R$ can distinguish various valley polarized and valley-coherent states. However, spin-polarized STM and magnetic exchange force microscopy experiments can precisely obtain the spin structures of the ground states~\cite{spinSTM-review}.

Although we here only accounted for the long-range Coulomb and the short-range backscattering interactions, as they are typically the most important contributions in a pristine system, our analysis can be generalized to explore some additional possibilities. For example, lattice-scale effects and electron-phonon interactions can support more exotic ground states, such as nematic (Pomeranchuk-type), stripe, translational symmetry breaking density-wave orders. 
{In addition, our theory can also be extended to study a topological insulator thin film, which resembles the conventional double-layer quantum Hall system in GaAs~\cite{Moon:1995ck}, where a plethora of broken symmetry phases associated with the layer degrees of freedom, such as a canted antiferromagnetic state~\cite{DasSarma1997:QHF,Zheng1997:QHF,DasSarma1998:QHF}, can be realized. We leave this topic for THBs for a future investigation.}

Even though in this work we completely focused on uniform ground states, the excitations above the ground state can carry a nontrivial topological structure, such as the \emph{valley skyrmions}. It is generally expected~\cite{Yang2006:QHFM} that at integer Landau level fillings, skyrmions can compete with simple particle-hole pairs for the lowest-energy charged excitations. If the system is placed slightly away from the integer filling factors, the skyrmions may also constitute the ground-state configuration in this system. Therefore, we expect that in the vicinity of $B_c^{(1)}$ (the tricritical points in Fig.~\ref{Fig:PhaseCoulomb}), where two valley-polarized phases and a valley-coherent phases meet, valley skyrmions can become energetically favorable. However, due to the anisotropy of the cyclotron orbits the energetic analysis of skyrmion excitations is an involved task, and deserves a separate quantitative investigation.

Finally we comment on the role of disorder. Notice that the $\bG$-polarized state does not break any discrete symmetry on the surface and hence is expected to be robust against disorder. In contrast, the $\bX$- or $\bY$-polarized, and $\bG$-$\bX$ or $\bG$-$\bY$ coherent states break the four-fold rotational symmetry, hence any generic point-like impurity will couple as \emph{random field} with electronic density when either $\bX$ or $\bY$ valley is occupied (fully or partially)~\cite{Imry-Ma}. Such random field coupling will lead to a proliferation of domain walls of opposite valley polarization/coherence at finite temperatures and destroy true long range ordering. However, the quantum Hall experiments are typically carried out at few hundred millikelvins, where the formation of domain walls can be energetically costly. In addition, the application of uniform strain can reduce the strength of such random field coupling, since it lifts the valley-Ising symmetry explicitly, providing stability to $\bX$-polarized and $\bG$-$\bX$ coherent phases against point-like impurities. It is worth pointing out that when $\bG$ and $\bX$, $\bY$ valleys carry opposite spin-momentum locking, the valley coherent phase carries a net magnetic moment, and time-reversal invariance preserving point-like impurities cannot couple with the ferrimagnet order parameter as random field. Thus, we expect that on a sufficiently clean surface of THBs, where residual electron-electron interaction is significantly strong, various phases we propose in this work can be observed in the quantum Hall regime, at least at low temperatures.

% Before concluding, we emphasize a few salient aspects of our theory and its consequences. First, our work focuses on various symmetry breaking quantum Hall phases at integer Landau level filling, where the interaction effects are substantially enhanced, leading to the manifestation of a plethora of spin- and valley-symmetry-breaking ground states. Furthermore, the Hartree-Fock theory is essentially exact (except for the Landau level coupling which can be small in the high-field limit) for integer Landau level filling. Although our analysis, at least in its simplest form, can not be extended for fractional fillings, some signatures of symmetry breaking can also be found in the fractional quantum Hall states. Finally, one may wonder whether various interaction-driven spin- and valley-polarized states can be realized even in the absence of magnetic fields. Such scenario is possibly unlikely with pure long-range Coulomb interaction, since in a two-dimensional Dirac system the kinetic and the Coulomb energy scale identically with the carrier density, thus not allowing any simple exchange-driven Bloch or Stoner type instabilities. On the other hand, in the presence of strong short-range interactions, these broken-symmetry ground states can, in principle, be realized~\cite{Roy:2014vv}. Such outcomes must be tested by suitable non-perturbative lattice Monte Carlo techniques, which, however, suffer from the sign problem, making precise quantitative predictions more challenging. In contrast, our analysis is controlled due to the large Landau level gap in the quantum Hall regime.  

\acknowledgements

This work was supported by JQI-NSF-PFC and LPS-MPO-CMTC. B.~R. thanks Jay D. Sau for valuable discussions, and Max Planck Institute for Complex System, Dresden during the workshop ``Quantum Design" (2015) where part of the manuscript was finalized. 

% \appendix

% \section{Zeeman coupling and Landau level crossing \label{Appendix:Zeeman}}

% In this Appendix, we show that inclusion of Zeeman coupling does not modify the critical field ($B_\ast$) above which there is no crossing between zeroth Landau levels and the rests. For simplicity we assume that applied magnetic field is exactly along the surface normal direction, so that there is no Zeeman coupling from its in-plane component. In this case, $B_\ast$ can be estimated from the following  
% \begin{align}
% \Delta_0 - \sqrt{\epsilon_0^2 B_\ast + (g_0\Ez B_\ast)^2} = g_1\Ez B_{\ast}. 
% \end{align}
% The general solution of $B_c$ is lengthy. But, valuable insights can be gained by studying some extreme limits. (i) When $g_1 = 0$~\cite{Zhang:2013jc}, the critical $B_\ast$ field gets reduced and given by
% \begin{align}
% B_\ast (g_1=0) = \dfrac{2\Delta_0^2}{\epsilon_0^2+\sqrt{\epsilon_0^4+4\Delta_0^2 g_0^2 \Ez^2}}. \label{Eq:Bc0}
% \end{align}
% (ii) In contrast, when $g_1 = g_0$ we have 
% \begin{align}
% B_\ast (g_1=g_0) = \dfrac{\Delta_0^2}{\epsilon_0^2 + 2g_0 \Delta_0 \Ez}. \label{Eq:BcEqual}
% \end{align}
% Hence, Zeeman coupling only reduces the critical field.

% Note that in the above two expressions no strain effects have been considered, because we believe strain is a weaker perturbation. For sufficiently large strain fields, however, these critical field estimations should be modified by replacing $\Delta_0$ by $\Delta_0 \pm \dst$. 

\bibliography{Bib-SmB6}
\end{document}